%% file: main.tex
\documentclass[a4paper,11pt]{article}
\pdfoutput=1 

\usepackage{jheppub} 
\usepackage{bm,amssymb,slashed,graphicx,multirow,soul,mathtools,xspace,array}  
\usepackage{float} 
\allowdisplaybreaks
\usepackage{bbold}
\usepackage{caption}
\usepackage{subcaption}

\newcommand{\gev}{{\rm GeV}}
\newcommand{\todo}[1]{{\color{red} \ifmmode\else[todo]\fi #1}}
\usepackage[usenames,dvipsnames]{xcolor}
     \definecolor{hgreen}{rgb}{0,.3,0}
      \definecolor{darkgreen}{rgb}{0.3,.8,0.2}
     \definecolor{hred}{rgb}{.3,0,0}
     \definecolor{hblue}{rgb}{0,0,.3}
     \definecolor{LightGray}{gray}{0.95}

\usepackage{slashed}

\usepackage{arydshln}
\usepackage{listings}
\definecolor{codegreen}{rgb}{0,0.6,0}
\definecolor{codegray}{rgb}{0.5,0.5,0.5}
\definecolor{codepurple}{rgb}{0.58,0,0.82}
\definecolor{backcolour}{rgb}{0.95,0.95,0.92}

\lstdefinestyle{pycode}{
  backgroundcolor=\color{backcolour},   commentstyle=\color{codegreen},
  keywordstyle=\color{magenta},
  numberstyle=\tiny\color{codegray},
  stringstyle=\color{codepurple},
  basicstyle=\ttfamily\footnotesize,
  breakatwhitespace=false,         
  breaklines=true,                 
  captionpos=b,                    
  keepspaces=true,                 
  numbers=left,                    
  numbersep=5pt,                  
  showspaces=false,                
  showstringspaces=false,
  showtabs=false,                  
  tabsize=2
}
\graphicspath{ {./images/} }

\usepackage{listings}

\def\Xint#1{\mathchoice
   {\XXint\displaystyle\textstyle{#1}}%
   {\XXint\textstyle\scriptstyle{#1}}%
   {\XXint\scriptstyle\scriptscriptstyle{#1}}%
   {\XXint\scriptscriptstyle\scriptscriptstyle{#1}}%
   \!\int}
\def\XXint#1#2#3{{\setbox0=\hbox{$#1{#2#3}{\int}$}
     \vcenter{\hbox{$#2#3$}}\kern-.5\wd0}}

\def\dashint{\Xint-}

\newcommand{\nn}{\nonumber}

\newcommand{\re}[0]{\mathrm{Re}\,}
\newcommand{\im}[0]{\mathrm{Im}\,}

\newcommand{\beq}{\begin{equation} }
\newcommand{\eeq}{\end{equation}} 
\newcommand{\bi}{\begin{itemize} }
\newcommand{\ei}{\end{itemize} }

\definecolor{Red}{rgb}{1.,0.,0.}
\definecolor{Grn}{rgb}{0.,0.75,0.}
\definecolor{Blu}{rgb}{0.,0.,1.}

\newcommand{\mycomment}[1]{}

\allowdisplaybreaks  

\preprint{JLAB-THY-24-4075}
\title{\boldmath Higgs-mixed Light Scalar Precisely}
\title{\boldmath Hadronic Decays of a Higgs-mixed Scalar}

\author[1,2]{P. Blackstone,}
\author[1,2,3]{J. Tarr\'{u}s Castell\`{a},}
\author[1,2,4,5]{E. Passemar,}
\author[6]{J. Zupan,}

\affiliation[1]{Department of Physics, Indiana University, Bloomington, IN 47408, USA}
\affiliation[2]{Center for Exploration of Energy and Matter,
		Indiana University, Bloomington, IN 47408, USA}
\affiliation[3]{Departament de F\'\i sica Qu\`antica i Astrof\'\i sica and Institut de Ci\`encies del Cosmos, 
Universitat de Barcelona, Mart\'\i $\;$ i Franqu\`es 1, 08028 Barcelona, Catalonia, Spain}
 \affiliation[4]{Departament de F\'isica Te\`orica, IFIC, Universitat de Val\`encia - CSIC, Apt. Correus 22085,
E-46071 Val\`encia, Spain}
\affiliation[5]{Theory Center, Thomas Jefferson National
	Accelerator Facility, Newport News, VA 23606, USA}
\affiliation[6]{Department of Physics, University of Cincinnati, Cincinnati, Ohio 45221, USA}

\emailAdd{pblackst@iu.edu}
\emailAdd{jtarrus@icc.ub.edu}
\emailAdd{emilie.passemar@ific.uv.es}
\emailAdd{zupanje@ucmail.uc.edu}

\abstract{One of the portals to new physics is a light scalar coupled to the Standard Model (SM) Higgs. In this paper we focus on hadronic decays of such a scalar in the regime where QCD dynamics is nonperturbative, resulting, e.g., in decays to pairs of pions or kaons, while also allowing for scalar couplings to the SM fermions to deviate from the Higgs-mixed light scalar limit. Representations of the corresponding form factors can be obtained using dispersive techniques, however, several sources of uncertainty affect the final results. We reexamine these decays, paying special attention to the quantification of uncertainties. For the light Higgs-mixed scalar scenario, we compare our results with previous works. For a general set of couplings of the light scalar to Standard Model fields, we provide a public code, {\tt hipsofcobra}, to compute the decay widths.
}

\begin{document} 

\maketitle

\flushbottom

\input{1_intro}
\input{2_shdff} 
\input{3_MO_problem} 
\input{4_matching}

\input{5_results}
\input{6_conclusion}

\appendix
\input{app_ff_expressions}

\bibliographystyle{JHEP}
\bibliography{BIB}

\end{document}

%% file: 1_intro.tex
\section{Introduction}

There are only a finite number of portals of lowest mass dimension between the Standard Model (SM) and light new physics: a light scalar, a light pseudoscalar (axion-like particle (ALP)), a spin 1/2 fermion (heavy neutral lepton), and a spin 1 vector boson (a dark photon) \cite{Patt:2006fw,Beacham:2019nyx}. These mediators have renormalizable couplings to the SM fields, apart from an axion, where the couplings start at dimension 5. It is thus not a stretch of the imagination to believe that such couplings may be the least suppressed ones mediating between the SM and the hidden sector.  Even though this is a relatively small set of models, each of them still has a number of freely adjustable couplings. For instance,  the flavor structure of mixings between heavy neutral leptons and the SM neutrinos can  significantly change the phenomenology, as can different choices for couplings of the ALP, or the presence of couplings of dark photon that are not due to the kinetic mixing. 

An exception to the rule is the light scalar portal \cite{Beacham:2019nyx},  for which a common benchmark is to take the limit of a  light Higgs-mixed scalar  $\phi$. In this limit  a light scalar mixes with the SM Higgs either through trilinear or quartic couplings, so that at low energies all the couplings to the SM fermions, gluons and photons are then fixed in terms of just one parameter, the mixing angle $\theta_h$. However, this is certainly not the most general possibility~\cite{Batell:2018fqo,Delaunay:2024}. Especially for the couplings of $\phi$ to the light SM fermions, $e, \mu, u, d, s$, which are suppressed by small Yukawa couplings $y_f\sim 10^{-5}-10^{-3}$, it is quite possible that higher dimensional operators may give larger contributions than the mixing with the Higgs (see, e.g., \cite{Altmannshofer:2015esa,Bishara:2015cha,Delaunay:2022grr,Altmannshofer:2018bch,Dery:2017axi,Kamenik:2023hvi,Altmannshofer:2016zrn} for examples).  

The decays of a light scalar with a mass between the two-pion threshold and $\sim2$~GeV into SM particles are expected to be dominated by pairs of pions and kaons. The computation of these decay widths is challenging, since Chiral Perturbation Theory ($\chi$PT), the effective field theory describing meson dynamics at low energy, is not applicable in the whole energy range, and in fact, the chiral expansion performs worse than one would naively expect due to the presence of scalar resonances in these channels. On the other hand, perturbative QCD is also not applicable in this range. As a result, dispersive techniques relying on unitarity and analyticity properties of amplitudes have emerged as the most useful tool to obtain a representation of the form factors associated with these decays. A key input in these representations are the pion and kaon scattering phase shifts. Early works used model phase shifts~\cite{Raby:1988qf,Truong:1989my}, while parametrizations fitted to experimental data have been used in Refs.~\cite{Donoghue:1990xh,Celis:2013xja,Winkler:2018qyg} (for data driven estimates of interactions of light vector and pseudoscalar particles see \cite{Baruch:2022esd,Aloni:2018vki,Ilten:2018crw}). 

In this manuscript we revisit the predictions for partial decay widths of such a light scalar, decaying into pairs of pions and kaons, while leaving the structure of couplings general. Since the flavor-violating couplings are constrained enough to not be relevant for $\phi$ decays, although they can be very important in $\phi$ production, we only need to focus on flavor-conserving couplings. We pay special attention at quantifying the errors on the predictions, and also provide a simple-to-use public code {\tt hipsofcobra} (\underline{Hi}ggs-\underline{P}ortal \underline{S}calar \underline{O}ff-\underline{F}lavor \underline{Co}upling \underline{B}ranching \underline{Ra}tios) to produce the decay widths in the general case. 

The paper is structured as follows. In Section~\ref{sec:shdff} we introduce the Lagrangian of a light scalar $\phi$ coupled to the SM fields, define the form factor for its decays into pairs of pions and kaons and its dispersive representation. In Section~\ref{sec:mop} we detail how to obtain the first piece of the dispersive representation, the Omnès function matrix, and Section~\ref{sec:matching} focuses on  the second piece, the determination of the subtraction polynomials. Our results for the form factors are presented in Section~\ref{sec:results}. We present our conclusions in Section~\ref{sec:conc}. In Appendix~\ref{app:code} we present the code accompanying this paper, that allows to compute the decay widths of light scalar for any value of the couplings to the SM fields. Finally, in Appendix~\ref{app:ff_expressions}, we collect the NLO $\chi$PT expressions used in the matching of the subtraction polynomials.

%% file: 2_shdff.tex
\section{Scalar hadronic decays form factor}\label{sec:shdff}

The effective Lagrangian of a light scalar $\phi$ coupled to the SM fields is given by~\cite{Carmi:2012in}
\beq
\begin{split}
	\mathcal{L}_{\rm eff} &=- \sum_q c_q\frac{m_q}{v_W}\bar{q}q\phi -
       \sum_\ell c_\ell\frac{m_\ell}{v_W}\bar{\ell}\ell \phi +
        c_g\frac{\alpha_s}{12\pi v_W} \phi G^a_{\mu\nu}G^{a\mu\nu}
         + c_\gamma\frac{\alpha}{\pi v_W} \phi F_{\mu\nu}F^{\mu\nu}, 
\end{split} \label{eq:lagrangian_general}
\eeq
where $v_W=246$\,GeV is the electroweak vev. The summation is over the quark flavors $q=u,d,s,c,b$ and charged leptons $\ell=e,\mu, \tau$, and we have assumed that flavor violation is negligible. 
For a Higgs-mixed scalar the couplings to quarks, gluons and leptons are given simply by
\beq
c_q = c_\ell = c_g = \sin \theta_h,
\label{eq:higgs_mixed_condition}
\eeq
where $\theta_h$ parametrizes the mixing between the scalar $\phi$ and the Standard Model Higgs. Further generalizations to the Lagrangian in Eq.~\eqref{eq:lagrangian_general}, such as enhanced coupling to neutrinos or dark matter,
are omitted from our analysis but can be straightforwardly included. 

Our objective is the determination of the following matrix element of a quark and gluon scalar current, which couples to the scalar in Eq.~\eqref{eq:lagrangian_general}, 
\begin{align}
G_P(s)=\langle P^+(p_1)P^-(p_2)|G|0\rangle\,,\quad P=\pi,K\,,\label{s4e6}
\end{align}
with 
\begin{align}
G=c_g\frac{\alpha_s}{12\pi v_W}G^{a}_{\mu\nu}G^{a\mu\nu}-\sum_q c_q\frac{m_q}{v_W}\bar{q}q\,,\label{s4e2}
\end{align}
and $s=(p_1+p_2)^2$, with $p_{1,2}$ the four momenta of the two outgoing mesons.

The form factors are analytic functions in the complex plane except on the cut on the positive real axis for $s>s_0$. 
$s_0$ represents the threshold, the two pion cut in our case: $s_0 \equiv 4 m_\pi^2$. 
From Cauchy's integral formula and Schwartz’s reflection principle we can find that
\begin{align}
G_P(s)=\frac{1}{\pi}\int^\infty_{s_0}\frac{dz}{z-s}\im G_P(z)\,,\label{dis_rel}
\end{align}
with $P=\pi,\,K\,$. For the remainder of this paper it should be understood that any equation depending on the subindex $P$ is valid both for pions and kaons.

The optical theorem, which can be derived from the unitarity of the $S$-matrix, tells us that the imaginary part of an amplitude is generated by the sum of all the intermediate state particles. Here, we are only interested in the hadronic contributions to the scalar currents with zero isospin and in particular to the ones that appear at small $s$. Thus we are only going to consider pairs of pions and kaons. Given these assumptions we can write the imaginary part as
\begin{align}
&\im\left[n_P G_P(s)\right]=\sum_{P'=\pi,K}(T^{*}(s))_{PP'}\sigma_{P'}(s)n_{P'}G_{P'}(s)\theta(s-4m^2_{P'})\,,\label{opt_theo}
\end{align}
where $n_\pi=\sqrt{3/2}$ and $n_K=\sqrt{2}$ are factors resulting from the projection of the pion and kaon states into isospin $I=0$, and
\begin{align}
\sigma_P(s)=\sqrt{1-\frac{4m^2_P}{s}}\,,\quad P=\pi,\,K\,.
\end{align}
The $S$-wave, isospin $I=0$, $T$-matrix is given by
\begin{align}
\bm{T}(s)\equiv \bm{T}^0_0(s) =\left(\begin{array}{cc}\frac{\eta^0_0(s)e^{2i\delta^0_0(s)}-1}{2i\sigma_{\pi}(s)} & |g^0_0(s)|e^{i\phi^0_0(s)} \\ |g^0_0(s)|e^{i\phi^0_0(s)} & \frac{\eta^0_0(s)e^{2i(\phi^0_0(s)-\delta^0_0(s))}-1}{2i\sigma_{K}(s)}\end{array} \right)\,.\label{t_matrix}
\end{align}
The three inputs of the $T$-matrix are: the $S$-wave, isoscalar, $\pi\pi$ phase shift $\delta^0_0(s)$ and the modulus, $|g^0_0|$, and phase, $\phi^0_0(s)$, of the $S$-wave isoscalar $\pi\pi\to K\bar{K}$ amplitude. The inelasticity $\eta^0_0(s)$ is related to $|g^0_0|$ by
\begin{align}
\eta^0_0(s)=\sqrt{1-4|g^0_0(s)|^2\sigma_{\pi}(s)\sigma_K(s)\theta(s-4m^2_K)}\,.
\end{align}

Introducing Eq.~\eqref{opt_theo} in Eq.~\eqref{dis_rel} we arrive to the following set of coupled integral equations
\begin{align}
n_PG_P(s)=\frac{1}{\pi}\sum_{P'=\pi,K}\int^\infty_{4m^2_\pi}\frac{dz}{z-s}(T^{*}(z))_{PP'}\sigma_{P'}(z)n_{P'}G_{P'}(z)\theta(z-4m^2_{P'})\,.\label{c_int_eq}
\end{align}

Our objective is to find a functional form of the form factors that fulfills Eq.~\eqref{c_int_eq}. To solve this problem it is additionally assumed that the form factors are  analytic in the complex $s$-plane, except on the cuts, and are real on the real $s$-axis below the cuts. This is the so-called coupled channel Muskhelishvili-Omn\`es problem~\cite{mushi,Omnes:1958hv}. The general solution is given by
\begin{align}
n_P G_P(s)=\Omega_{PP'}(s)Q_{G_{P'}}(s)\,,\label{gen_sol}
\end{align}
where the sum over $P'=\pi,K$ is implied. The $\bm{\Omega}$ matrix encodes the two independent canonical solutions arranged as columns
\begin{align}
\bm{\Omega}(s)=\left(\begin{array}{cc} \Omega_{\pi\pi}(s) & \Omega_{\pi K}(s) \\ \Omega_{K \pi}(s) & \Omega_{KK}(s) \end{array}\right)\,,\label{omega_def}
\end{align}
and $\bm{Q}_{G}(s)=(Q_{G_{\pi}},\,Q_{G_{K}})^T \!$ are known as the subtraction polynomials.

We show how to obtain $\bm{\Omega}(s)$ and $\bm{Q}_{G}(s)$, including a detailed uncertainty analysis, in sections~\ref{sec:mop} and \ref{sec:matching}, respectively. We present our results for $ G_P(s)$ in section~\ref{sec:results}.

%% file: 3_MO_problem.tex
\section{Muskelishvili-Omn\`es problem} \label{sec:mop}

The matrix $\bm{\Omega}$ satisfies the following set of coupled Muskhelishvili-Omn\`es singular integral equations
\begin{align}
\bm{\Omega}(s)=\frac{1}{\pi}\int^{\infty}_{4m^2_{\pi}}\frac{dz}{z-s}\left(\bm{T}(z)\right)^{*}\bm{\Sigma}(z)\bm{\Omega}(z)\,,\label{omega_int_eq}
\end{align}
with $\bm{\Sigma}(s)=\text{diag}(\sigma_\pi(s)\theta(s-4m^2_{\pi}),\sigma_K(s)\theta(s-4m^2_{K}))$. The two independent solutions are generated choosing the normalization $\bm{\Omega}(0)=\mathbb{1}$.

Let us discuss under which asymptotic conditions, $s\to\infty$, a solution to Eq.~\eqref{omega_int_eq} exists. If we take the determinant of the equation for the discontinuity of the form factor across the cut, the matrix equation reduces to a one-dimensional equation for which an analytical solution is available~\cite{mushi,Omnes:1958hv}. The asymptotic behavior can then be obtained~\cite{Moussallam:1999aq}:
\begin{align}
\text{det}(\bm{\Omega})\stackrel{s\to\infty}{\sim}s^{-\text{Arg}(\text{det}(S))/\pi},
 \label{asyn_limit}
\end{align}
with $S$ the $S$-matrix associated to the $T$-matrix in Eq.~\eqref{t_matrix}. Assuming that the off-diagonal terms of $S$ vanish in the asymptotic limit, then $\text{Arg}(\text{det}(S))$ is just the sum of the asymptotic behaviors of the eigen phase shifts. Since each component of $\bm{\Omega}$ must vanish at least as $1/s$~\cite{Lepage:1980fj}, Eq.~\eqref{asyn_limit} establishes a constraint on the asymptotic behavior of the $T$-matrix for solutions of integral equations to exist:
\begin{align}
\lim_{s\to\infty}\text{Arg}(\text{det}(S(s)))\geq 2\pi\,,
\end{align}
with $2$ corresponding to the pion and kaon channels considered in the present work.

In order to solve Eq.~\eqref{omega_int_eq} it is convenient to split the real and imaginary parts of $\bm{\Omega}(s)$. Using Sokhotsky's formula we find
\begin{align}
\re \bm{\Omega}(s) &= \frac{1}{\pi} \dashint^{\infty}_{4m^2_{\pi}}\frac{dz}{z-s}\im \bm{\Omega}(z)\,,\label{re_pv}\\
\im \bm{\Omega}(s)  &= \left(\bm{T}(s)\right)^{*}\bm{\Sigma}(s)\bm{\Omega}(s)\,,
\end{align}
with the dashed integral representing the Cauchy principle value. Using the fact that $\im  \bm{\Omega}(s)$ is itself a real number one can find that
\begin{align}
\im \bm{\Omega}(s) =   \mathbf{X}(s) \re \bm{\Omega}(s)\,,\label{im_re}
\end{align}
where we have defined
\begin{align}
\mathbf{X}(s) &= i\left(\mathbb{1}-\frac{\mathbb{1}}{\mathbb{1} - i \bm{T}(s)^* \bm{\Sigma}(s) }\right) .
\end{align}

\noindent Using Eq.~\eqref{im_re} on Eq.~\eqref{re_pv} we arrive at the following integral equation for the real part of $\bm{\Omega}$:
\begin{align}
\re \bm{\Omega}(s) &= \frac{1}{\pi} \dashint^{\infty}_{4m^2_{\pi}}\frac{dz}{z-s}\mathbf{X}(z) \re \bm{\Omega}(z)\,.\label{re_omega_int_eq}
\end{align}
Now we need to solve Eq.~\eqref{re_omega_int_eq} to obtain the real part of $\bm{\Omega}$, and then use Eq.~\eqref{im_re} to find the imaginary one, thus obtaining the complete solution.

\subsection{Numerical Solution}
The system of integral equations is solved by using the method presented in Refs.~\cite{Moussallam:1999aq,Descotes-Genon:2000pfd} and we follow the implementation of Ref.~\cite{TarrusCastella:2021pld}. The method consists in a first step to approximate Eq.~\eqref{re_omega_int_eq} by a linear system of equations obtained by discretising the integral. The second step is to find a solution of the linear system that fulfills the normalisation condition of the Omnès functions $\bm{\Omega}(0)=\mathbb{1}$. An alternative method based on an iterative solution was used in Refs.~\cite{Donoghue:1990xh,Celis:2013xja}.

We begin by dividing the integration domain $[4m^2_\pi,\infty)$ into $M$ segments. Let $a_0=4m^2_\pi$ and $a_1, \ldots , a_{M-1}$ be the rest of the boundaries of the segments such that the first segment is $[a_0,a_1]$, the second $[a_1,a_2]$, etc. The last segment goes up to infinity $[a_{M-1},\infty)$. We break up the integral in Eq.~\eqref{re_omega_int_eq} into several pieces corresponding to each of the $M$ segments of the integration domain we have just defined. For each piece we rescale the integration variable such that the integration range becomes $[-1,1]$:
\begin{align}
I_m(s) &= \dashint_{a_{m-1}}^{a_m} dz \frac{\mathbf{X}(z) \re \bm{\Omega}(z)}{ z - s } = \dashint_{-1}^1 dy \frac{g_m(y)}{y-y_m(s)}\,,\quad m=1,..,M-1\,,\label{int_seg} 
\end{align}
with
\begin{align}
y_m(s)&=(2s-a_m-a_{m-1})/(a_m-a_{m-1})\,,\\
z_m(y)&=[(a_m-a_{m-1})y+a_m+a_{m-1}]/2\,,\\
g_m(y)&=\mathbf{X}(z_m(y)) \re \bm{\Omega}(z_m(y))\,.
\end{align}
Similarly the last segment is rescaled as
\begin{align}
I_M(s) &= \dashint_{a_{M-1}}^\infty dz \frac{\mathbf{X}(z) \re \bm{\Omega}(z)}{z-s} = \dashint_{-1}^1 dy \frac{g_M(y)}{y-y_M(s)}\,,\label{int_last_seg} 
\end{align}
with 
\begin{align}
y_{M}(s)&=1-2a_{M-1}/s\,,\\
z_{M}(y)&=2a_{M-1}/(1-y)\,,\\
g_M(y)&=(z_{M}(y)/s)\mathbf{X}(z_M(y)) \re \bm{\Omega}(z_M(y))\,.
\end{align}
Next we  write $g(y)$ in the basis of Legendre polynomials
\begin{align}
g_m(y) = \sum_{k=0}^\infty c^{(m)}_k P_k (y), \quad {\rm with} \quad c^{(m)}_k = \frac{2k+1}{2} \int_{-1}^1 dy \, g_m(y) P_k(y).
\end{align}
This allows the exact integration of the principal value using the formula
\begin{align}
\dashint_{-1}^1 dy \frac{P_k(y)}{y-y_m(s)}=-2 \re Q_k(y_m(s))\,,\label{pv_int}
\end{align}
where $Q_k(x)$ is a Legendre function of the second kind.\footnote{Note that the formula in Eq.~\eqref{pv_int} is a special case of a more general version in terms of the full integral in the left-hand side and the full (i.e., not just the real part) $Q_k$ in the right-hand side.}

The coefficients of the Legendre polynomial expansion, $c^{(m)}_k$, are approximated using the Gauss-Legendre quadrature~\cite{Abramowitz}
\begin{align}
c^{(m)}_k &\approx \frac{2k+1}{2} \sum_{n=1}^{N} w_n g_m(y_n) P_k(y_n)\,,
\end{align}
where  $y_n$ are the roots of the $N$-th Legendre polynomial, $P_N(y_n)=0$, and $w_n$ are the so-called weights 
\begin{align}
w_n = \frac{2}{1-(y_n)^2 \left[P_N' (y_n) \right]^2}\,.
\end{align}
The approximation would be exact, if $g_i$ were a degree $N$ polynomial.

Furthermore, the Legendre polynomial expansion of $g_i$ is also truncated to a degree $N$ polynomial. The specific value of $N$ chosen is arbitrary, however, larger values are expected to lead to a better approximation. Therefore we arrive at the following expression for the integrals in Eqs.~\eqref{int_seg} and \eqref{int_last_seg}   
\begin{align}
I_m(s) &\approx \sum_{n=1}^N g_m(y_n) W_n(y_i(s))\,,\label{int_seg_approx}
\end{align}
with
\begin{align}
W_n(y)=-w_n \Bigg[\sum_{k=0}^{N-1} (2k+1) P_k(y_n) \re Q_k(y) \Bigg]\,.\label{bWn}
\end{align}
Note that, since  $P_N(y_n)=0$, cutting the sum in Eq.~\eqref{bWn} at $N-1$ is equivalent to cutting it at $N$. Additionally it can be shown that
\begin{align}
W_n(y) = -w_n\left[\frac{ 1 - (N+1) P_{N+1}(y_n) \re Q_N(y) }{ y - y_n}\right]\,.\label{bWnr}
\end{align}
The expression in Eq.~\eqref{bWnr} is faster to compute, in particular if the values of the Legendre function of the second kind are to be computed with high accuracy, as is necessary in our case. However, as we will see in the following, we will need to evaluate $W_n(y_n)$, in which case the denominator in Eq.~\eqref{bWnr} goes to zero. In this case, using Eq.~\eqref{bWn} might be more practical than a careful numerical computation of  $\lim_{y\to y_n}W_n(y)$.

Using Eq.~\eqref{int_seg_approx} in the coupled integral equation~\eqref{c_int_eq} we find the following approximated expression for the real part of $\bm{\Omega}$
\begin{align}
 \pi \re \bm{\Omega}(s) = \sum_{m=1}^{M-1}&\sum_{n=1}^{N}W_n(y_m(s))\bm{X}(s^{(m)}_n) \re\bm{\Omega}(s^{(m)}_n)\nn \\
 +\frac{1}{s}&\sum_{n=1}^{N}s^{(M)}_nW_n(y_M(s))\bm{X}(s^{(M)}_n) \re\bm{\Omega}(s^{(M)}_n)\,,\label{c_int_dis}
\end{align}
with $s^{(m)}_n=z_m(y_n)$, i.e., the set of values of $s$ resulting from the inverse mapping for each segment of the $n$ roots of the Legendre polynomial. Using Eq.~\eqref{c_int_dis} we can compute any value of $ \re \bm{\Omega}(s)$ from a finite set of values $\re \bm{\Omega}(s^{(m)}_n)$. Thus, the next step is to obtain these values. To do so, we evaluate Eq.~\eqref{c_int_dis} at each $s^{(m)}_n$, creating a linear system of equations for $\re \bm{\Omega}(s^{(m)}_n)$. 
Let us note that each column of $\re \bm{\Omega}(s)$, say $(\re \Omega_1, \re \Omega_2)^T$, corresponds to a solution of Eq.~\eqref{re_omega_int_eq}, therefore we only need to consider one of them in order to set up the linear system. One can choose any way to arrange the set of values $\left\{(\re \Omega_1(s^{(m)}_n),\re \Omega_2(s^{(m)}_n))^T\right\}$ into a vector, in our case we have used
\begin{align}
\bm{v}_{\Omega}=\left(\re \Omega_{1}(s_1^{(1)}), \cdots, \re \Omega_{1}(s_1^{(2)}), \cdots, \re \Omega_{1}(s_N^{(M)}), \re \Omega_{2}(s_1^{(1)}),\cdots, \re \Omega_{2} (s_M^{(N)})\right)^T\,,
\end{align}
that is, stacking them in order of increasing $s^{(m)}_n$ values. The length of $\bm{v}_{\Omega}$ vector is $2MN$. Similarly, one can construct a $(2N\!M)\!\times\! (2N\!M)$ matrix, $\bm{\mathcal{M}}$, such that Eq.~\eqref{c_int_dis} becomes
$\bm{\mathcal{M}}\bm{v}_{\Omega}=\bm{0}$. Strictly speaking, this is a homogeneous system with nontrivial solutions only when $\det(\bm{\mathcal{M}})=0$. However, in practice, due to the approximations we have carried out and the finite computational accuracy, the determinant doesn't exactly vanish. Rather than seeking the precise null-space of $\bm{\mathcal{M}}$, it is more convenient to consider the normalization constraint $\bm{\Omega}(0) = \mathbb{1}$, which adds two extra non-homogeneous equations to the linear system. Note that there are two sets of extra equations, corresponding to the two columns on both sides of the normalization constraint, which generate the two independent solutions of the Muskhelishvili-Omn\`es  problem. One can use any of these two sets of extra equations to replace two of the equations in the homogeneous system to obtain a system with one unique solution, however, this leads to a numerically unstable solution. It turns out to be more practical to add the normalization equations to the homogeneous system creating a $(2N\!M+2)\!\times\! (2N\!M)$ overdetermined non-homogeneous system. 

An approximate solution of this system of equations can be obtained using the Moore-Penrose pseudoinverse~\cite{penrose_1955}. Let us denote the overdetermined system as $\widetilde{\bm{\mathcal{M}}}\tilde{\bm{v}}_\Omega=\bm{b}$, where the tilde denotes the addition of the normalization equations and $\bm{b}$ is a vector of zeros except in the entry corresponding to the normalization equation. The singular value decomposition of a matrix $\bm{A}$ is given by $\bm{A}=\bm{U}\bm{D}\bm{V}^T$, where $\bm{U}$ and $\bm{V}$ are orthogonal matrices whose columns are left- and right-singular vectors, respectively. These correspond to the eigenvectors of $\bm{A}\bm{A}^T$  and $\bm{A}^T\bm{A}$. $\bm{D}$ is a non-square diagonal matrix, whose elements are known as the singular values of $\bm{A}$. The pseudoinverse can be computed in terms of the elements of this decomposition as $\bm{A}^+=\bm{V}\bm{D}^+\bm{U}^T$, where $\bm{D}^+$ is obtained by taking the reciprocal of the non-zero elements of $\bm{D}$ and then taking the transpose of the resulting matrix. Finally, the solution $\tilde{\bm{v}}_\Omega=\bm{\widetilde{\mathcal{M}}}^+\bm{b}$ is the one that minimises $\lVert \widetilde{\bm{\mathcal{M}}}\tilde{\bm{v}}_\Omega-\bm{b}\rVert$, i.e., the least squares solution.

\subsection{Omn\`{e}s function results}\label{omnes_res}

\begin{figure}[ht!]
    \centering
    \begin{subfigure}{0.49\textwidth}
        \centering
        \includegraphics[width=\textwidth]{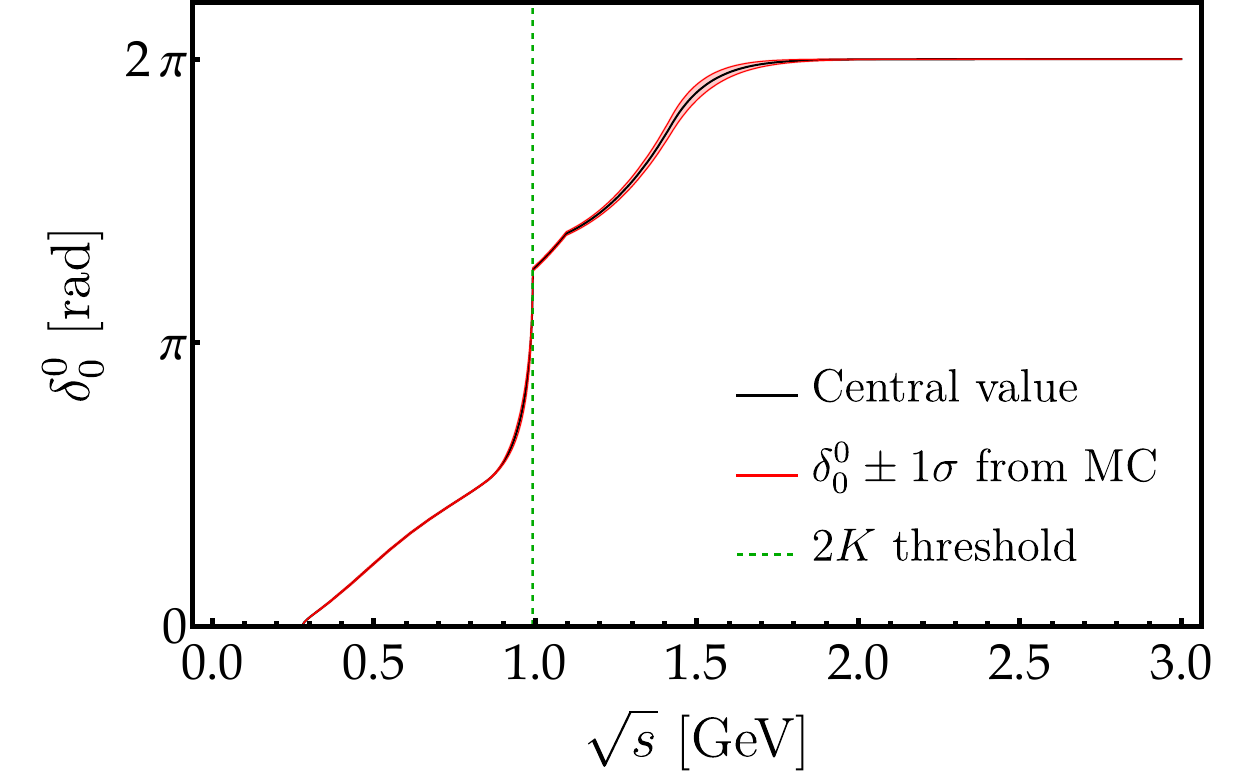}
        \caption{$\delta^0_0$, the $\pi\pi \to \pi\pi$, $\ell=0$ scattering phase shift. \hspace{2cm} }
        \label{subfig:delta1_plot}
    \end{subfigure}
    \hfill
    \begin{subfigure}{0.49\textwidth}
        \centering
        \includegraphics[width=\textwidth]{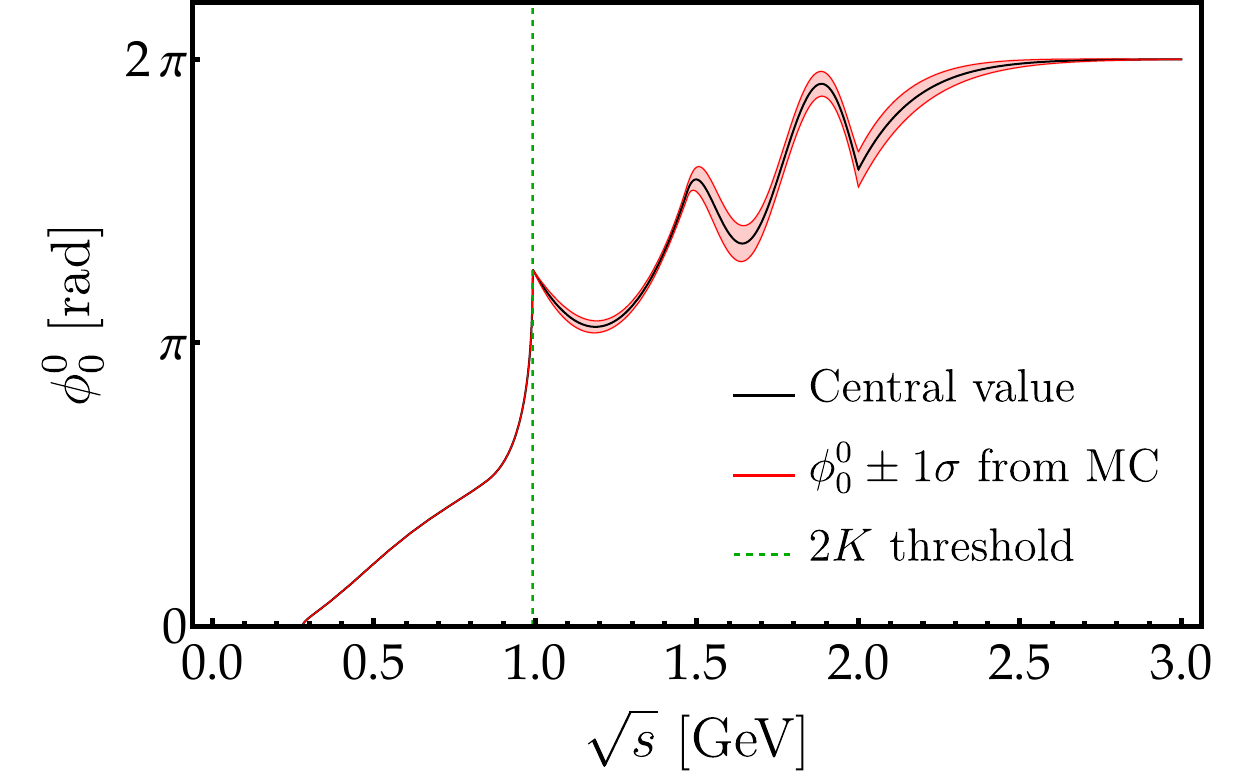}
        \caption{$\phi^0_0$, the $\pi\pi \to KK$, $\ell=0$ scattering phase shift.} 
        \label{delta2_plot}
    \end{subfigure}
    \hfill
    \begin{subfigure}{0.49\textwidth}
        \centering
        \includegraphics[width=\textwidth]{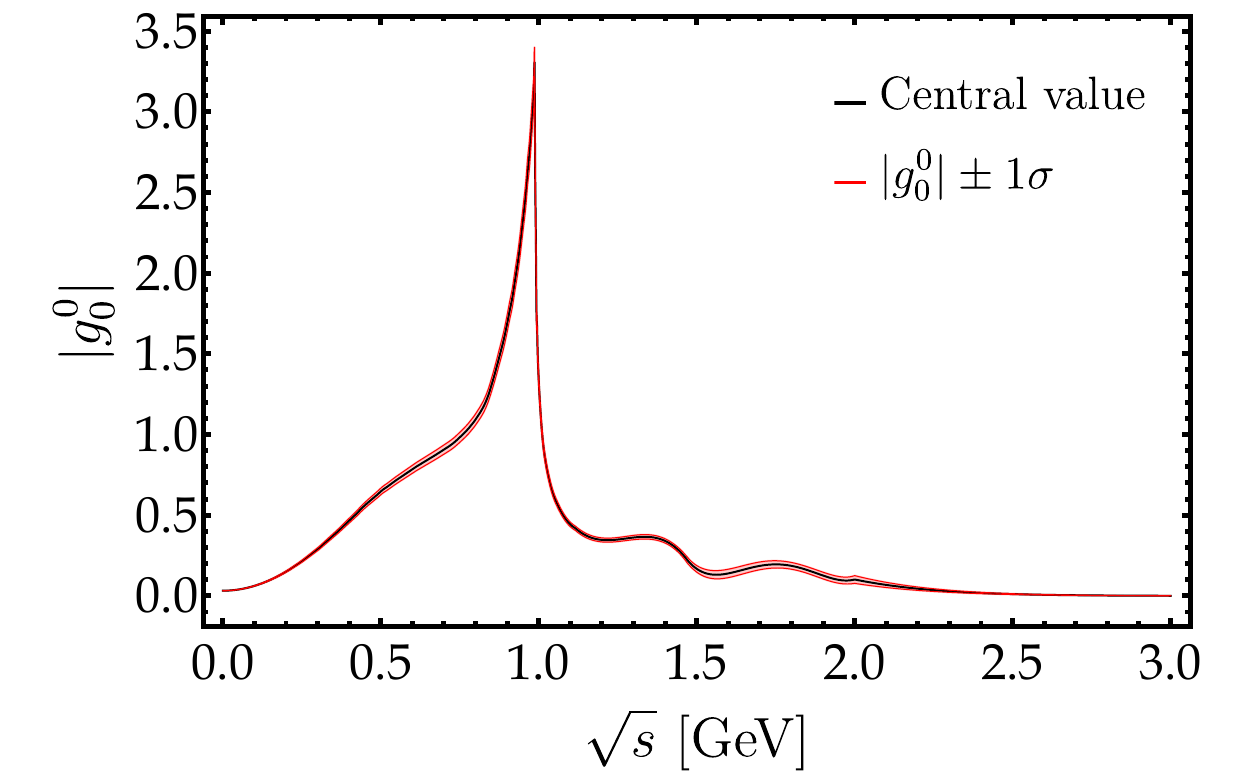}
        \caption{$|g_0^0|$, the magnitude of $\pi\pi \to KK$ amplitude. \hspace{2cm} }
        \label{subfig:g00_plot}
    \end{subfigure}
    \hfill
    \begin{subfigure}{0.49\textwidth}
        \centering
        \includegraphics[width=\textwidth]{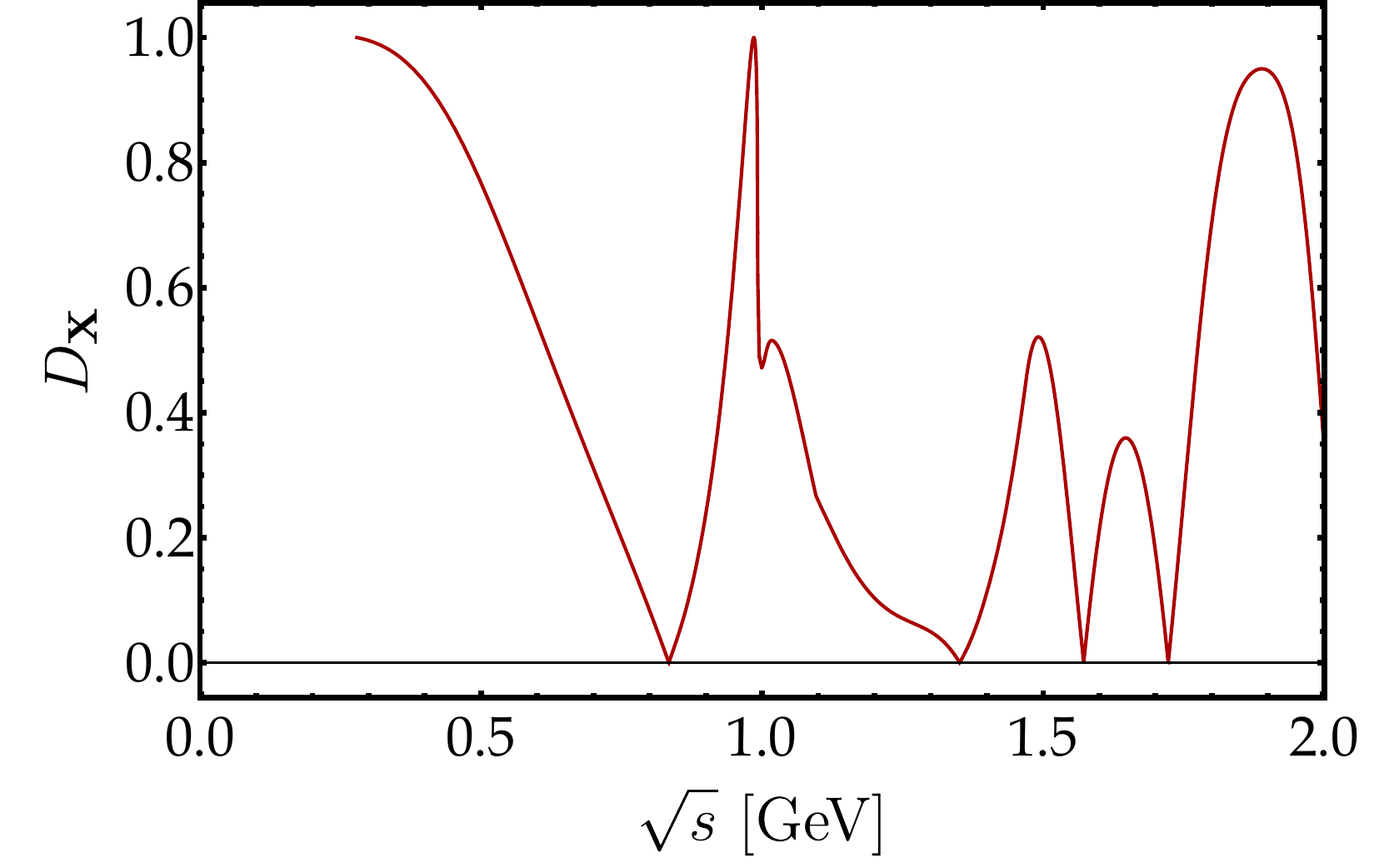}
        \caption{A particular instance of $D_{\mathbf{X}}(s)$, the roots of which are taken as division boundaries.}
        \label{subfig:xdenom_plot}
    \end{subfigure}
    \caption{Phenomenological input for the $S$ matrix. Error bounds are evaluated by Monte Carlo, iterating the parameter selection many times to establish variance.}
    \label{fig:pheno_input}
\end{figure}

In order to solve the Muskhelishvili-Omn\`es problem we need to input the $S$-wave isoscalar $\pi\pi \to \pi\pi$ phase shift $\delta^0_0(s)$ and the modulus, $|g^0_0|$, and phase, $\phi^0_0(s)$, of the $S$-wave isoscalar $\pi\pi\to K\bar{K}$ amplitude, which determine the $T$-matrix in Eq.~\eqref{t_matrix}. We use the parametrizations of Ref.~\cite{Garcia-Martin:2011iqs} with the ``CFD'' parameter set for $\delta^0_0(s)$ and of Ref.~\cite{Pelaez:2020gnd}  with the ``CFD$_\text{c}$'' parameter set for $|g^0_0|$ and $\phi^0_0(s)$. These parametrizations are obtained from fits to experimental data and include uncertainty values for the parameters, which reflect the uncertainties on the experimental data. To take this into account, we consider a Gaussian distribution for each parameter with the mean and standard deviations corresponding to the parameter values and uncertainties from Refs.~\cite{Garcia-Martin:2011iqs,Pelaez:2020gnd} and randomly sample these to obtain $100$ sets of parameters. From each parameter set we obtain phase shift input. In Fig.~\ref{fig:pheno_input} we plot profiles of these inputs, where the central line and bands correspond to the average and standard deviation of the input values for such a set of iterations at each $s$.

The experimental data used in Refs.~\cite{Garcia-Martin:2011iqs,Pelaez:2020gnd} to fit the parametrizations of the phase shifts extends only up to $\sqrt{s}=1.42$~GeV for $\delta^0_0(s)$ and $\sqrt{s}=2$~GeV for $|g^0_0|$ and $\phi^0_0(s)$. However, to solve the Muskhelishvili-Omn\`es problem we need parametrizations up to $s\to \infty$. To obtain these, we continue the parametrizations of Refs.~\cite{Garcia-Martin:2011iqs,Pelaez:2020gnd} from their values, $\alpha_{\rm cut}$, at the cut-off, $s_{\rm cut}$, to their asymptotic value, $\alpha_{\infty}$, using a function of the form
\begin{align}
    f(s) = \alpha_\infty + (\alpha_{\rm cut} - \alpha_\infty) \exp\!\big(\! - A ( s - s_{\rm cut} ) \big)\,.
\end{align}
The free parameter $A>0$ tunes the slope of the exponential at $s_{\rm cut}$. In the case of $\delta^0_0$ the slope is fixed to keep differentiability. For $\phi_0^0$ and $|g_0^0|$ the slope must change sign in order to reach the asymptotic values and differentiability is not maintained. These asymptotic values are $2\pi$ for the phases $\delta^0_0(s)$ and $\phi^0_0(s)$ and $0$ for $|g^0_0|$. This continuation is carried out for each of the 100 parametrizations of the phase shifts resulting from randomly sampling the parameter distributions. Note that we do not attempt to provide a measure of the uncertainty associated to these continuations to the asymptotic values.
As a result, the uncertainty bands for the phase shifts in Fig.~\ref{fig:pheno_input} tend to zero as these approach their asymptotic values.

Finally, before presenting our results, we state how we divide the integral into subintervals. First, we choose the kaon threshold, $4m_K^2$, as one of the boundaries and the last segment to start at  $a_{M-1}= 4 \, \gev^2$. The rest of the boundaries are set at the roots of
\begin{align}
D_{\mathbf{X}}(s) &= \left| \det \left( \mathbb{1} - i \bm{T}^* \bm{\Sigma}\right) \right|,
 \label{eq:dx_function}
\end{align}
which are natural choices for division boundaries. Fig.~\ref{subfig:xdenom_plot} shows $D_{\mathbf{X}}$ for a representative selection of input parameters. These generally drift from iteration to iteration and so must be updated each time. 

\begin{figure}[ht!]
    \centering
    \begin{subfigure}{0.45\textwidth}
        \centering
        \includegraphics[width=\textwidth]{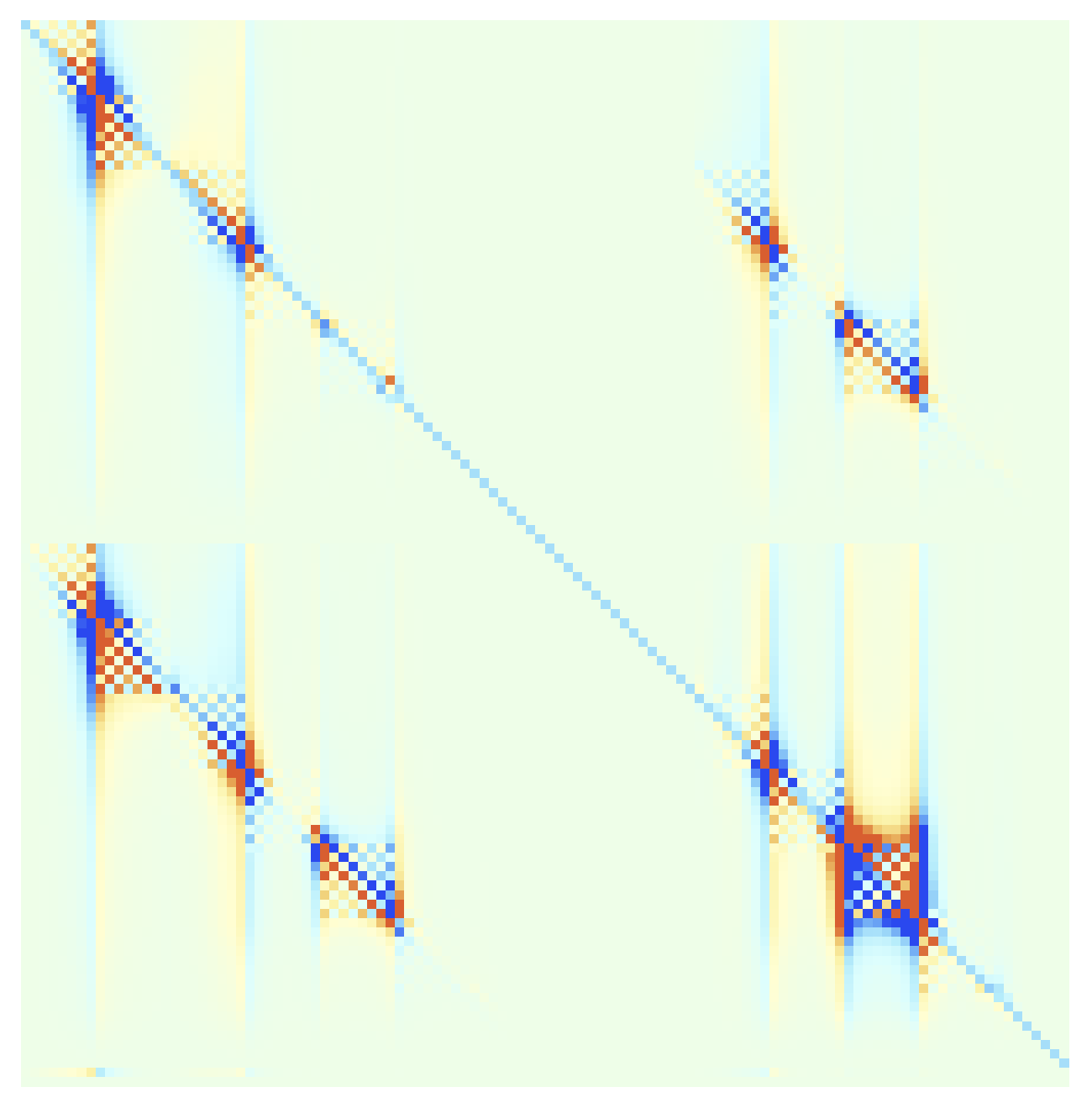}
        \caption{}
        \label{subfig:matrix_image_deg_8}
    \end{subfigure}
    \hfill
    \begin{subfigure}{0.45\textwidth}
        \centering
        \includegraphics[width=\textwidth]{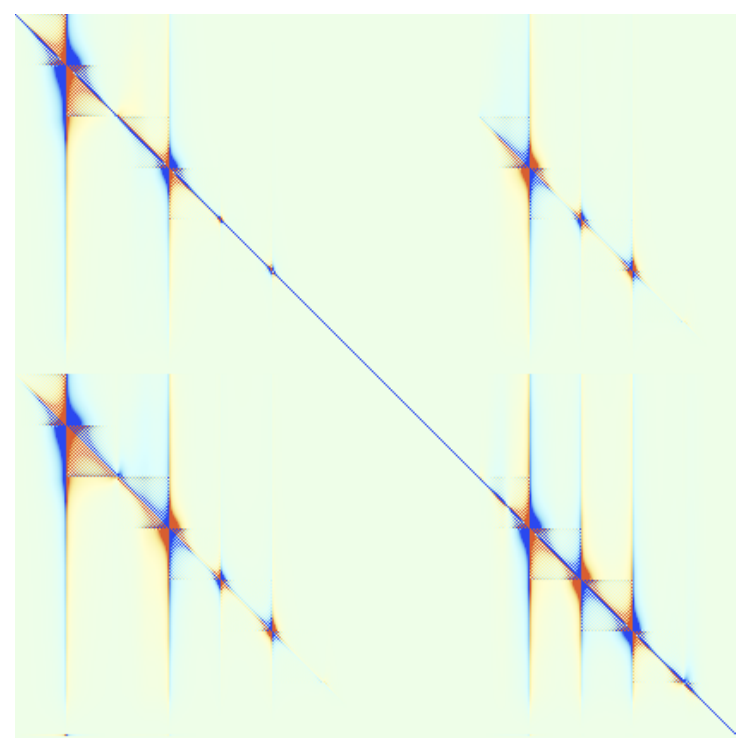}
        \caption{}
        \label{subfig:matrix_image_deg_30}
    \end{subfigure}
    \caption{Image representations of the matrix $\mathcal{M}$ for (a) $N=8$ and (b) $N=30$, using a representative instance of the input discussed in Sec.~\ref{omnes_res}. In both (a) and (b), the full range $[4m_\pi^2,\infty)$ has been broken up into $7$ intervals. Red (blue) pixels represent positive (negative) entries, while increasing saturation indicates absolute value of the matrix elements. The color scaling omits outliers outside of the $0.01$ to $0.99$ quartile range to prevent them from skewing the representation.}
    \label{fig:matrix_images}
\end{figure}

We illustrate the structure of the matrix $\bm{\mathcal{M}}$ in Fig.~\ref{fig:matrix_images}. The matrix elements are shown as red or blue pixels for positive or negative values, respectively, and values close to zero are less saturated. The figures are shown for different resolutions of the approximation, $N=8$ and $N=30$. The key observation is that the main contributions to each row come from elements near the diagonal. This can be understood as follows: the mapping of $s^{(m)}_n$ to the rescaled segment variable $y_{m'}(s^{(m)}_n)$ produces values outside the interval $[-1,1]$ if $m'\neq m$. Since the values of $\re Q_k(y)$ decrease very rapidly with $y$ for $|y|>1$ the corresponding contributions to Eq.~\eqref{c_int_dis} become negligible. On the other hand the contributions with $m'=m$ are the most important. In this case we have that $y_{m}(s^{(m)}_n)=y_n$ and therefore the elements near the diagonal correspond to $W_{n'}(y_n)$, which can be computed once, saving computational time, and used for all the entries with $m'=m$. Moreover, the exactly diagonal elements include $W_{n}(y_n)$, which, as we discussed, should be evaluated carefully if computed using Eq.~\eqref{bWnr} or instead by using Eq.~\eqref{bWn}. An interesting consequence of this discussion is that the arbitrary continuation of the phase shifts from the maximum values of $s$ for which we have experimental data to the asymptotic values has a suppressed effect on the results for the Omnès functions in the range of $s$ for which we have experimental data.

\begin{figure}[ht!]
    \centering
    \begin{subfigure}{0.49\textwidth}
        \centering
        \includegraphics[width=\textwidth]{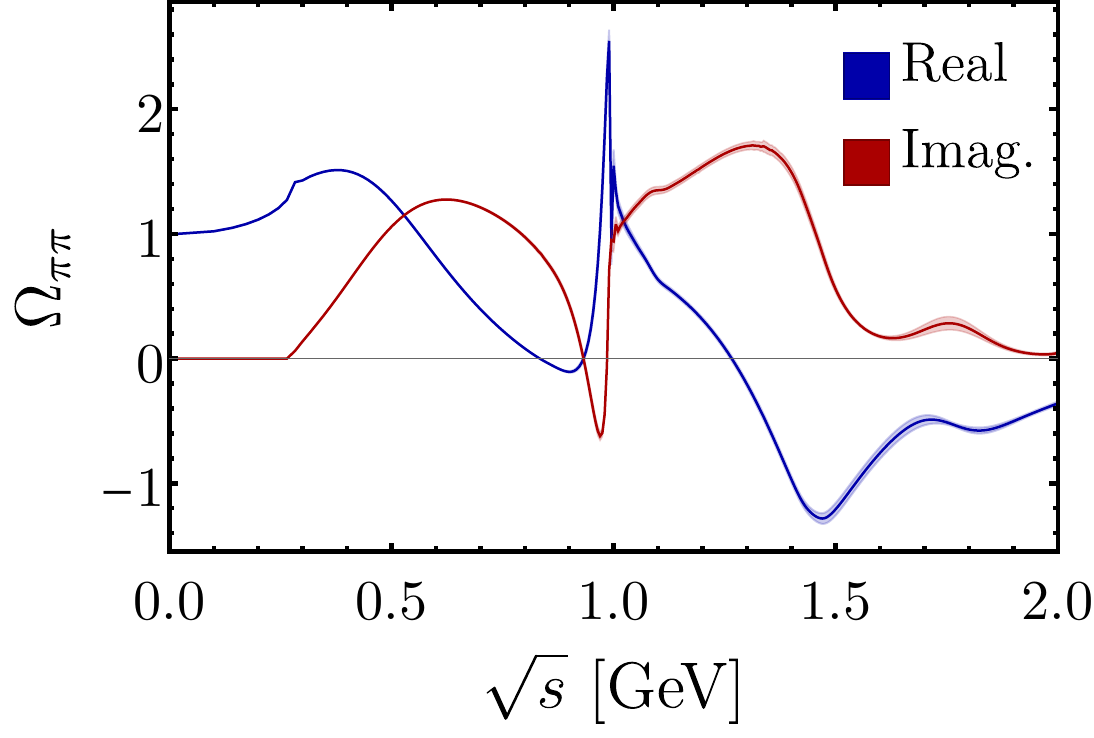}
        \caption{}
        \label{subfig:c1_plot}
    \end{subfigure}
    \hfill
    \begin{subfigure}{0.49\textwidth}
        \centering
        \includegraphics[width=\textwidth]{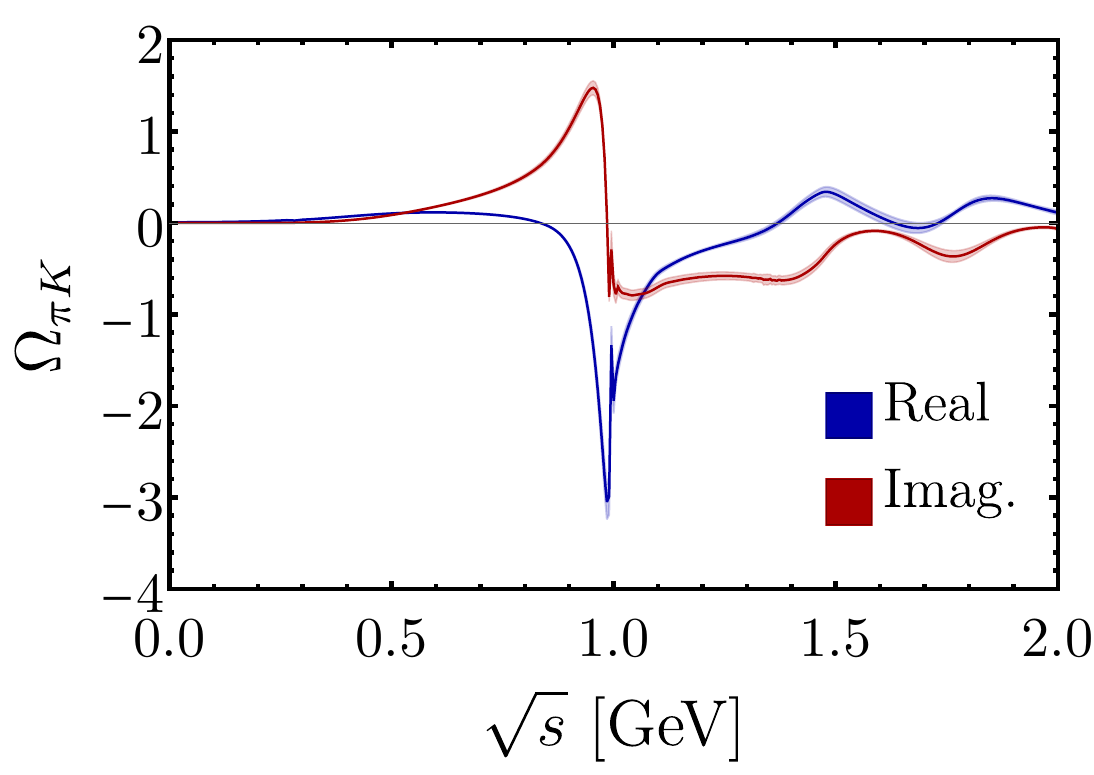}
        \caption{}
        \label{d1_plot}
    \end{subfigure} \\
    \centering
    \begin{subfigure}{0.49\textwidth}
        \centering
        \includegraphics[width=\textwidth]{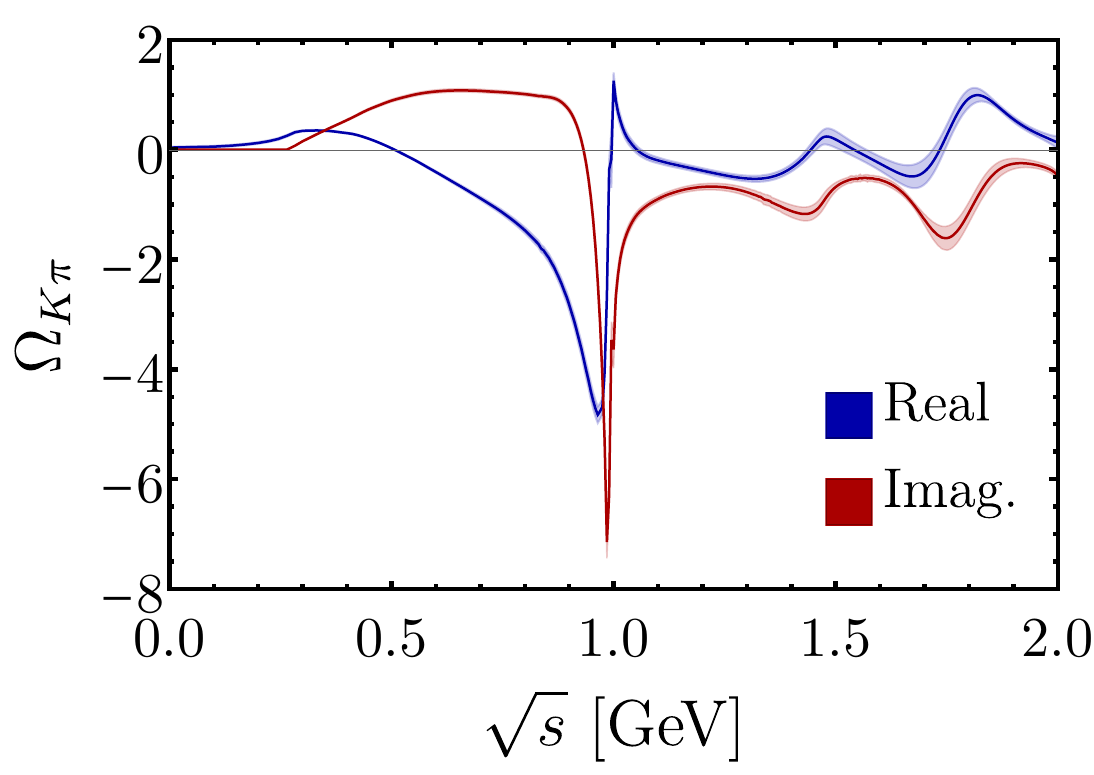}
        \caption{}
        \label{subfig:c2_plot}
    \end{subfigure}
    \hfill
    \begin{subfigure}{0.49\textwidth}
        \centering
        \includegraphics[width=\textwidth]{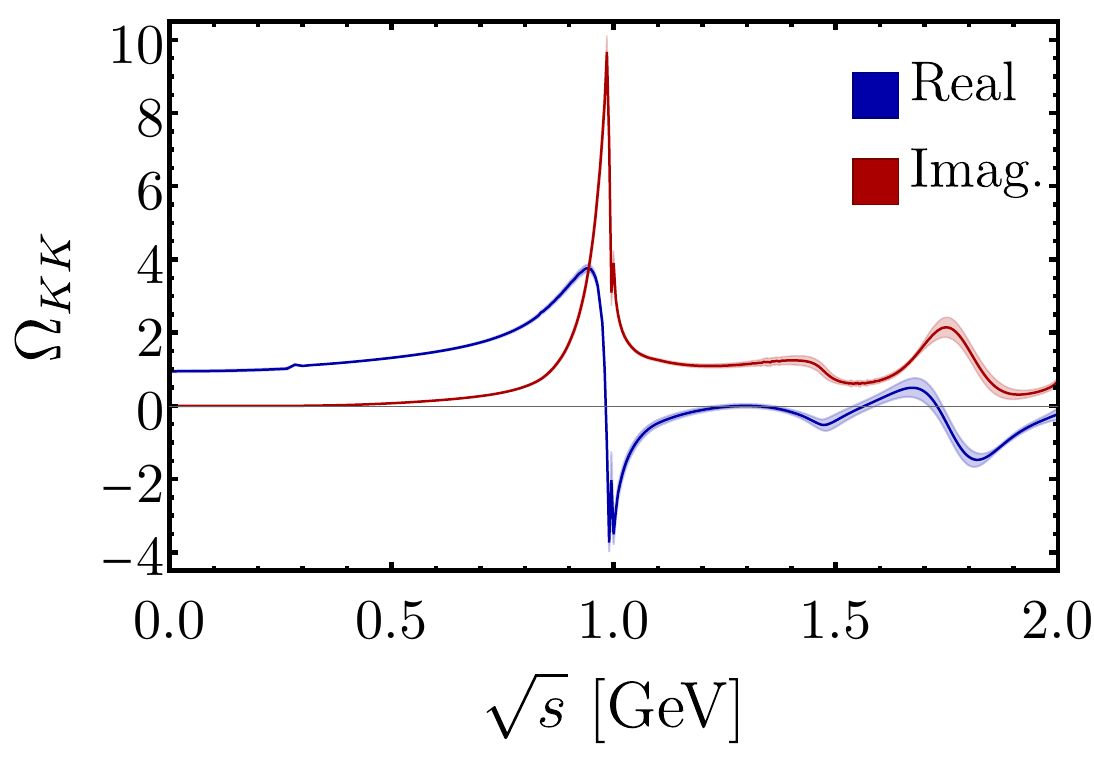}
        \caption{}
        \label{d2_plot}
    \end{subfigure}
    \caption{Omn\`{e}s functions with uncertainty bands generated from $100$ iterations at an approximation degree of $N=30$ over $7$ intervals. The $1\sigma$ uncertainty bands are due to varying the phase shifts and $|g_0^0|$.}
    \label{fig:FFs_with_uncertainty}
\end{figure}

We run the numerical solution algorithm, using $N=30$ and $M=7$ intervals\footnote{In most parametrizations, the roots of $D_\mathbf{X}$ are such that there are $7$ intervals. For some samplings there are fewer roots, and we instead choose extra divisions in order to keep the number consistent between iterations.} for each of the $100$ sets of parametrizations of the phase shifts that we have described above. The results for the Omnès functions are shown in Fig.~\ref{fig:FFs_with_uncertainty}. The central line and the uncertainty bands correspond to the average and standard deviation of the 100 iterations.\footnote{To plot the results from a single solution of the set of iterations one does not need to compute any extra values of $\re \bm{\Omega}(s)$ for $s>4m^2_\pi$ since the sampling of $s$ given by $\{s_n^{(m)}\}$ is already quite dense. However, the set of values $\{s_n^{(m)}\}$ does depend on the position of the boundaries of the segments and these in turn depend slightly on the parametrization of the phase shifts, and thus differ for each iteration. Therefore, one should recompute the values of $\re \bm{\Omega}(s)$  for some fixed sampling of $s$ for all iterations in order to find the mean and standard deviation on that sampling set. Most rigorously, one would do the computation for this fixed sampling of $s$ using Eq.~\eqref{c_int_dis}, however this is computationally expensive. Instead, we take advantage of the fact that the original samplings $\{s_n^{(m)}\}$ are already dense, to extrapolate the values for the new fixed sampling using linear interpolation.}

The results for the central line of the Omnès functions shown in Fig.~\ref{fig:FFs_with_uncertainty} are in full agreement with those of Ref.~\cite{TarrusCastella:2021pld} which were obtained using the same phase shift parametrizations from Refs.~\cite{Garcia-Martin:2011iqs,Pelaez:2020gnd} as in the present work, except for minor differences in the driving of the phase shifts to their asymptotic values, and the same numerical solution technique detailed in this section. In Ref.~\cite{Hoferichter:2012wf} the Omnès functions were computed using the phase shifts parametrizations from Refs.~\cite{Caprini:2011ky,Buettiker:2003pp} and using a numerical solution also based on Refs.~\cite{Moussallam:1999aq,Descotes-Genon:2000pfd}. Their result shows less pronounced peaks around the kaon threshold with smaller to no undulations for $\sqrt{s}=1.4-2$~GeV range. Moreover, $\re \Omega_{K\pi}$ does not show a peak with positive values after the kaon threshold. More recent results can be found in Ref.~\cite{Danilkin:2020pak}. These have been obtained using the $N/D$ method~\cite{Chew:1960iv}, up to $\sqrt{s}\sim 1.44$~GeV, and include uncertainty bands. The general structure, including the size of the error bands, matches our results. The main difference of Ref.~\cite{Danilkin:2020pak} with our results is less pronounced peaks around the kaon threshold. The origin of this difference could be tentatively attributed to a smoother parametrization of the phase shift in the kaon threshold region.

%% file: 4_matching.tex
\section{Low-energy matching}\label{sec:matching}

If the Omn\`es function matrix has been determined the second piece to obtain $G_P(s)$ is to determine the subtraction polynomials $Q_{G_P}(s)$. To do so, we will make use of the fact that for small $s$ the form factors $G_P(s)$ can be computed in $\chi$PT. Thus, the subtraction polynomials can be determined from a set of matching conditions between the chiral and dispersive representations of the form factor $G_P(s)$. Unfortunately, there is an ambiguity in the literature on the exact procedure to follow to perform the matching. In the following we will present two possible matching procedures and the corresponding results.

\subsection{Chiral representation}

Let us first obtain the chiral representation of $G_P(s)$. It is convenient to introduce the trace energy-momentum tensor
\begin{align}
\label{eq:theta:mu:mu}
\theta^{\mu}_{\mu}=\frac{\beta(\alpha_s)}{4\alpha_s}G^{a}_{\mu\nu}G^{a\mu\nu}+\sum_{q}(1-\gamma_m) m_q\bar{q}q\,,
\end{align}
where $\beta$ is the QCD beta function and $\gamma_m$ is the anomalous dimension of the quark operator.  If we rewrite $G$ in Eq.~\eqref{s4e2} in terms of the trace of the energy-momentum tensor we find
\begin{align}
G&=\xi_{g}\theta^{\mu}_{\mu}-\sum_{q}\left[\xi_q+(1-\gamma_m)\xi_{g}\right]m_q\bar{q}q\,,\label{s4e3}
\end{align}
where we have defined
\begin{align}
\xi_{g}&=\frac{c_g\alpha^2_s}{3\pi v_W\beta(\alpha_s)}\,, \qquad \xi_q=\frac{c_q}{v_W}\,.
\end{align}
Let us introduce the form factor definitions from Ref.~\cite{Donoghue:1990xh}:
\begin{align}
\theta_{P}(s)&=\langle P^+(p_1)P^-(p_2)|\theta^{\mu}_{\mu}|0\rangle\,,\label{s4e4}\\
\Gamma_{P}(s)&=\langle P^+(p_1)P^-(p_2)|\sum_{q=u,d}m_q\bar{q}q|0\rangle\,,\\
\Delta_{P}(s)&=\langle P^+(p_1)P^-(p_2)|m_s \bar{s}s|0\rangle\,.\label{s4e5}
\end{align}
Using Eqs.~\eqref{s4e3}-\eqref{s4e5} we can rewrite the form factor in Eq.~\eqref{s4e6} as follows:
\begin{align}
G_P(s)&=\xi_{g}\theta_{P}(s)-\left[\left(\hat{\xi}+(1-\gamma_m)\xi_{g}\right)\Gamma_P(s)+\left(\xi_s+(1-\gamma_m)\xi_{g}\right)\Delta_P(s)\right]\label{s4e14}\,,
\end{align}
with $\hat{\xi}=(\xi_u m_u+\xi_d m_d)/(m_u+m_d)$. Isospin-breaking terms are neglected, which should indeed be small unless $c_u$ and $c_d$ are orders of magnitude different. In our numerics we conservatively consider only the case $\xi_{u}=\xi_{d}$, i.e., we define a parameter
\beq
\label{eq:cud}
c_{ud}=c_u=c_d,
\eeq
and thus, equivalently, $\xi_{ud}=\xi_u=\xi_d$, while we leave the case of $c_u\ne c_d$ (including isospin breaking effects) for future work.

The three form factors in Eqs.~\eqref{s4e4}-\eqref{s4e5} can be evaluated at small $s$~\cite{Shifman:1978zn,Voloshin:1980zf,Novikov:1980fa,Chivukula:1989ds} using $\chi$PT except for a possible normalization. The leading order expressions~\cite{Bijnens:1998fm} for  $\Gamma_P(s)$ and $\Delta_P(s)$ are given by 
\begin{align}
\Gamma^{\rm \chi PT}_{P}(s)&=\Gamma^{\rm }_{P}(0)\left[1+\mathcal{O}(p^2)\right]\,,\label{s4e8a}\\
\Delta^{\rm \chi PT}_{P}(s)&=\Delta^{\rm }_{P}(0)\left[1+\mathcal{O}(p^2)\right]\,,\label{s4e8b}
\end{align}
with the normalization constant fixed by the Feynman-Hellman theorem
\begin{align}
\Gamma_{P}(0)&=\sum_{q=u,d} m_q\frac{\partial m^2_P}{\partial m_q} \,,\label{s4e9a}\\
\Delta_{P}(0)&=m_s\frac{\partial m^2_P}{\partial m_s}\,.\label{s4e9b}
\end{align}
Since the space integral of $\theta^{\mu 0}$ is the momentum operator, the normalization of the trace of the energy-momentum tensor matrix element is fixed by kinematic constraints. The leading order $\chi$PT expressions reads as
\begin{align}
\theta^{\rm \chi PT}_{P}(s)&=s+2m^2_{P}+\mathcal{O}(p^4)\,.\label{s4e10}
\end{align}
The pseudoscalar mass, $m^2_P$, in Eqs.~\eqref{s4e8a}, \eqref{s4e8b} and \eqref{s4e10} should, in principle, be evaluated at leading  order in $\chi$PT in order to be consistent with the chiral order of these expressions. However, the normalization expressions in Eqs.~\eqref{s4e9a} and \eqref{s4e9b} hold for the physical masses. Furthermore, kinematics also constrains
\begin{align}
\theta_{P}(0)&=2m^2_{P}\,,\label{s4e10b}
\end{align}
with $m^2_P$ at the physical mass. Therefore, at $s=0$ the form factors are related as follows:
\begin{align}
\Gamma_{P}(0)+\Delta_{P}(0)=\frac{1}{2}\theta_{P}(0)=m^2_P\,.
\label{s4e13}
\end{align}
Using Eqs.~\eqref{s4e8a}, \eqref{s4e8b}, \eqref{s4e10} and \eqref{s4e13} in Eq.~\eqref{s4e14} we can find the chiral representation of $G_P(s)$
\begin{align}
G^{\rm \chi PT}_P(s)&=\xi_g s+\left[(1+\gamma_m)\xi_g-\hat{\xi}\right]m^2_P+\left(\hat{\xi}-\xi_s\right)\Delta_{P}(0)+\mathcal{O}(p^4)\,. 
\end{align}
As before, the form factor at $s=0$ can be evaluated at the pseudoscalar physical mass
\begin{align}
G_P(0)&=\left[(1+\gamma_m)\xi_g-\hat{\xi}\right]m^2_P+\left(\hat{\xi}-\xi_s\right)\Delta_{P}(0)\,. \label{s4e33}
\end{align}

\subsection{Subtraction polynomial}

The construction of a dispersive representation for a scalar current such as to the one in Eq.~\eqref{s4e2} was carried out in Ref.~\cite{Donoghue:1990xh} by Donoghue, Gasser and Leutwyler (DGL).\footnote{The only difference is in the numerical factors multiplying the operators.} A dispersive representation of each of the matrix elements in Eqs.~\eqref{s4e4}-\eqref{s4e5} was built separately and then added together to form the dispersive representation of Eq.~\eqref{s4e6}. 

In principle, the degree of the subtraction polynomial must be chosen such that dispersive representation of the form factor does not diverge in the large $s$ limit. Since the Omn\`es functions have been constructed such that in this limit they behave as $s^{-1}$, the subtraction polynomial should be a constant. We can determine the subtraction constant by evaluating the form factor in Eq.~\eqref{gen_sol} at $s=0$. Recalling that the Omn\`es functions are normalized as $\Omega_{PP'}(0)=\delta_{PP'}$ we arrive at
\begin{align}
Q_{\Gamma_P} &=n_P \Gamma_P(0)\,,\\
Q_{\Delta_P} &=n_P \Delta_P(0)\,.
\end{align}

Unfortunately, if we apply the same procedure to $\theta_{P}$ we find results which are inconsistent with the small $s$ representation in Eq.~\eqref{s4e10}. Namely, if the subtraction polynomial is just a constant, the leading order dependence on $s$ in Eq.~\eqref{s4e10} cannot be reproduced by the small $s$ expansion of the dispersive representation. To reconcile these two representations for small $s$, it is necessary to let the subtraction polynomial to be of degree one. Using such a subtraction polynomial in Eq.~\eqref{gen_sol} leads to divergent form factors in the limit $s\to \infty$. We can make sense of this by noticing that the two channel set up discussed in Section~\ref{sec:mop} is only valid  up to values of $s$ for which new channels open. Moreover, the experimental data for the phase shifts is also only available up to certain energies. It can be shown, see section 5.2 of Ref.~\cite{Moussallam:2011zg}, that for $s$ below the one in which these effects appear, a linear term in the subtraction polynomial accounts for such high energy contributions. Therefore, we will proceed with a degree one subtraction polynomial and only use the dispersive representation of the form factors in the range of $s$ for which there is available experimental data for the phase shifts.

To determine the degree one subtraction polynomial $Q_{\theta_{P}}$ two matching conditions are needed. Let
\begin{align}
Q_{\theta_{P}}=Q^{(0)}_{\theta_{P}}+Q^{(1)}_{\theta_{P}}s\,,
\end{align}
The first condition is to match the value of the form factor at $s=0$ as we have done previously. This determines the constant terms of the polynomial
\begin{align}
Q^{(0)}_{\theta_{P}}&= n_P\theta_{P}(0)\,.\label{s4e11}
\end{align}
The second condition is to match the slope of the form factor also at $s=0$, which determines the coefficients of the linear terms of the subtraction polynomial to be
\begin{align}
Q^{(1)}_{\theta_{P}}&=n_P\dot{\theta}_P(0)-\dot{\Omega}_{PP}(0)Q^{(0)}_{\theta_{P}}-\dot{\Omega}_{PP'}(0)Q^{(0)}_{\theta_{P'}}\,,\label{s4e12}
\end{align}
where the dots indicate derivatives with respect to $s$. Using Eq.~\eqref{s4e11} in Eq.~\eqref{s4e12} we arrive at 
\begin{align}
Q^{(1)}_{\theta_{P}}&=n_P\left(\dot{\theta}_P(0)-\dot{\Omega}_{PP}(0)\theta_{P}(0)\right)-n_{P'}\dot{\Omega}_{PP'}(0)\theta_{P'}(0)\,.
\end{align}

Following the procedure proposed in Ref.~\cite{Donoghue:1990xh} by DGL, we then construct the dispersive representation of the form factor $G_P(s)$ by inserting the individual dispersive representations for each of the currents $\theta_P$, $\Gamma_P$ and $\Delta_P$ that we have just determined. This results in $G_P(s)$ given by Eq.~\eqref{gen_sol} with the following degree one subtraction polynomial
\begin{align}
Q_{G_{P}}= Q^{(0)}_{G_{P}}+Q^{(1)}_{G_{P}}s,
\end{align}
with
\begin{align}
\begin{split}
Q^{(0)}_{G_{P}}&=\xi_g Q^{(0)}_{\theta_{P}}-\left[\left(\hat{\xi}+(1-\gamma_m)\xi_{g}\right)Q_{\Gamma_P}+\left(\xi_s+(1-\gamma_m)\xi_{g}\right)Q_{\Delta_P}\right]
\\
&=n_P G_P(0)\,,\label{s4e15}
\end{split}
\\
Q^{(1)}_{G_{P}}&=\xi_g Q^{(1)}_{\theta_{P}}\,.\label{s4e16}
\end{align}
This set of matching polynomials for $G_P$ we refer to as the {\bf DGL} low-energy matching condition.

A different way of building the dispersive representation of $G_P(s)$ is to consider the whole form factor instead of splitting it into the constituent $\Gamma, \Delta, \theta$ components. In this case, we can determine the subtraction polynomial in an analogous way as we did for $\theta_P(s)$, using Eqs.~\eqref{s4e11} and \eqref{s4e12}, replacing $\theta_P(s)$ by $G_P(s)$. We find the following
\begin{align}
Q^{(0)}_{G_P} =&\, n_P G_P(0)\,,\label{s4e17}
\\
\begin{split}
Q^{(1)}_{G_P} =&\, n_P\dot{G}_P(0)-\dot{\Omega}_{PP}(0)Q^{(0)}_{G_{P}}-\dot{\Omega}_{PP'}(0)Q^{(0)}_{G_{P'}}
\\
=&\, n_P\left(\dot{G}_P(0)-\dot{\Omega}_{PP}(0) G_P(0)\right)-n_{P'}\dot{\Omega}_{PP'}(0) G_{P'}(0)\,.
\label{s4e18}
\end{split}
\end{align}
This set of polynomials we refer to as the {\bf BTPZ} matching condition. 

If we compare these with our previous results we can see that Eq.~\eqref{s4e17} matches the result in Eq.~\eqref{s4e15}. On the other hand, the coefficient of the linear term in Eq.~\eqref{s4e18} contains additional contributions that are not present in Eq.~\eqref{s4e16}, since the linear term in Eq.~\eqref{s4e18} depends on $Q_{\Gamma_P}$, $Q_{\Delta_P}$ and $Q^{(0)}_{\theta_P}$ (through the expression for $G_P(s)$), while in Eq.~\eqref{s4e16} it depends only on $Q^{(0)}_{\theta_P}$.

The origin of the discrepancy is that constructing a dispersive representation of $G_P(s)$ as a whole is equivalent to constructing it from the representations of $\Gamma_P, \Delta_P, \theta_P$ only if all the subtraction polynomials are of degree one and not just the one for $\theta_P$. The use of degree one subtraction polynomials for $\Gamma_P$ and $\Delta_P$ can be justified following the arguments of Ref.~\cite{Moussallam:2011zg} that we discussed for $\theta_P$, namely that the linear term accounts for inevitable missing high energy contributions to the dispersive integral. An additional way to sidestep this issue~\cite{BM} is to count terms of the type $m^2_P\dot{\Omega}_{PP'}(0)$ as ${\cal O}(p^2)\ll 1$, in which case the matching produces the same results for $G_P$ as a whole or as a sum of the $\Gamma_P, \Delta_P, \theta_P$ components as, even considering degree one subtraction polynomials for all the form factors, only $Q^{(1)}_\theta$ has leading order contribution. Note, however, that applying this counting implies that Eq.~\eqref{s4e12} reduces to $Q^{(1)}_{\theta_{P}}=n_P\dot{\theta}_P(0)$, and therefore produces different results for the dispersive representation of $\theta_P$ than the DGL~\cite{Donoghue:1990xh} ones.

\subsection{Uncertainty from low-energy matching}\label{sec:uncertainty_le_matching}

In the previous section we have determined the subtraction polynomials for the form factors in terms of their values at $s=0$, and for the case of $G_P$ and $\theta_P$, also the values of their slopes. The former can be obtained from the masses of the pseudoscalar mesons  and their derivatives with respect to the quark masses (see Eqs.~\eqref{s4e9a}, \eqref{s4e9b} and \eqref{s4e10b}). In order to do so we use the $\pi$ and $K$ masses at NLO in $\chi$PT, see Appendix~\ref{app:ff_expressions} for details. The estimate for the uncertainty of the form factors at $s=0$ is determined from the propagation of the uncertainty in the (combinations of) low-energy constants (LECs) present in the expressions of the masses reported in Ref.~\cite{Dowdall:2013rya}. Compared to these, the uncertainty associated to NNLO  $\chi$PT contributions, estimated from their parametric size, is much smaller for the pion form factors and smaller but of the same order for the kaon ones: ${(m_K / 4 \pi F_\pi )^4 \sim 6 \% }$.

At leading order $\dot{\theta}_P(0)=1$. The NLO order expressions can be found in Ref.~\cite{Donoghue:1991qv}. However, to carry out this computation, it is required to extend the standard NLO $\chi$PT Lagrangian to include the metric as an external field, which introduces a set of additional LECs. The only estimation of the additional LECs relevant for the evaluation of $\dot{\theta}(0)$ that we are aware of is from Ref.~\cite{Donoghue:1991qv} itself and is obtained from a dispersive representation of $\theta_P(s)$ similar to the one explored in this work and therefore cannot be used. Thus, we set $\dot{\theta}_P(0)$ to its leading order value and estimate the uncertainty as the parametric size of the NLO contributions given by $( m_\pi / 4 \pi F_\pi )^2 \approx 0.02$ and $( m_K / 4 \pi F_\pi )^2 \approx 0.24$ for the $\pi$ and $K$ form factors, respectively. We can improve on the determination of $\dot{\theta}_K(0)$ following the suggestion by DGL~\cite{Donoghue:1990xh} to relate the slope terms of $\theta_\pi$ and $\theta_K$ by devising a sum rule for the difference $\theta_K - \theta_\pi$. The sum rule is constructed by imposing that this difference tends to zero as $s\to\infty$ as it is an $SU(3)$ breaking effect. In this way, we find $\dot{\theta}_K(0)=1.17$. The propagation of the uncertainty of $\dot{\theta}_\pi(0)$, from the parametric estimate, through the sum rule leads to a negligible uncertainty for $\dot{\theta}_K(0)$ compared to its own parametric estimate, as the canonical solutions $\Omega_{ij}$ are quite precise in our analysis. We take the conservative approach and take the parametric estimate as the uncertainty of $\dot{\theta}_K(0)$.

Finally, the value of $G_P(0)$ is obtained using Eq.~\eqref{s4e33} and $\dot{G}_P(0)=\xi_g \dot{\theta}_P(0)$. For these two quantities, and in general for the computation of $G_P(s)$, we use $\beta(\alpha_s)$ and $\gamma_m$ truncated at the $\alpha_s^{4}$ order  from Refs~\cite{vanRitbergen:1997va} and \cite{Chetyrkin:1997dh}, respectively. This provides a significant improvement from the traditional truncation at leading order employed in Refs.~\cite{Voloshin:1980zf,Novikov:1980fa,Chivukula:1989ds,Donoghue:1990xh,Celis:2013xja,Winkler:2018qyg,Pineda:2019mhw,TarrusCastella:2021pld}. The uncertainty associated with this truncation is negligible compared to the other sources of uncertainty. We present a summary of the form factor values at $s=0$, as well as the uncertainties,  in Tab.~\ref{tab:ff_values_at_zero}. When combining the $6\%$ uncertainty from the NNLO parametric estimate in quadrature with the LEC contributions, it is only $\Delta_K, \theta_K$, and $G_K$ with the Higgs-mixed scalar couplings whose uncertainties are affected, mainly as a consequence of reporting the uncertainty of those to two digits. 

\begin{table}[t!]
\begin{center}
\begin{tabular}{ c  l } 
 \hline\hline
 $F_P$ & Value at $s=0$ \\
 \hline
 $\Gamma_\pi$ & $(0.98 \pm 0.01) m_\pi^2$
  \\  
 $\Gamma_K$   & $(1.5 \pm 0.2) \frac{1}{2} m_\pi^2$
 \\ 
 $\Delta_\pi$ & $(0.009 \pm 0.008) \left( m_K^2 - \frac{1}{2} m_\pi^2 \right)$
  \\  
 $\Delta_K$   & $(1.14 \pm 0.18) \left( m_K^2 - \frac{1}{2} m_\pi^2 \right) $
 \\  
 $\theta_\pi$ & $(0.98 \pm 0.01) 2m_\pi^2$ 
 \\  
 $\theta_K$   & $(1.16\pm 0.14)2m_K^2 $
 \\  
 $G_\pi$      & $(1.07 \pm 0.05) \left( - \frac{11}{9} \frac{m_\pi^2}{v_W} \right)$
  \\  
 $G_K$        & $(1.21 \pm 0.14) \left( - \frac{11}{9}  \frac{m_K^2}{v_W}  \right) $ \\

 $\dot{\theta}_\pi$ & $1.00\pm 0.02$ \\  
 $\dot{\theta}_K$ & $1.17\pm 0.28$ \\ 
 $\dot{G}_\pi$ & $\left(1.9 \pm 0.1\right)\left( - \frac{2}{9v_W} \right) $ \\  
 $\dot{G}_K$ & $\left(2.4 \pm 0.9\right)\left( - \frac{2}{9v_W} \right)$ \\ 
 
 \hline  \hline
\end{tabular}
\end{center}
\caption{Values of form factors at $s=0$ from the analysis described in Section~\ref{sec:uncertainty_le_matching}. The mass-containing expressions that the numerical factors multiply match the form of the LO results in $\chi$PT and $\alpha_s$ expansion, where here $m_\pi$ and $m_K$ are the physical masses. The exception is $\Delta_\pi$, for which the corresponding prediction is $0$ and is written in units of $m_K^2 - \frac{1}{2} m_\pi^2$ simply for comparison. The values for $G_P(0)$ and $\dot{G}_P(0)$ listed here correspond to Higgs-mixed scalar case with $c_{ud}=c_s=c_g=1$.
}
\label{tab:ff_values_at_zero}
\end{table}

%% file: 5_results.tex
\section{Results}\label{sec:results}
\subsection{Form factors}

\begin{figure}[ht!]
     \centering
     \begin{subfigure}{0.49\textwidth}
         \centering
         \includegraphics[width=\textwidth]{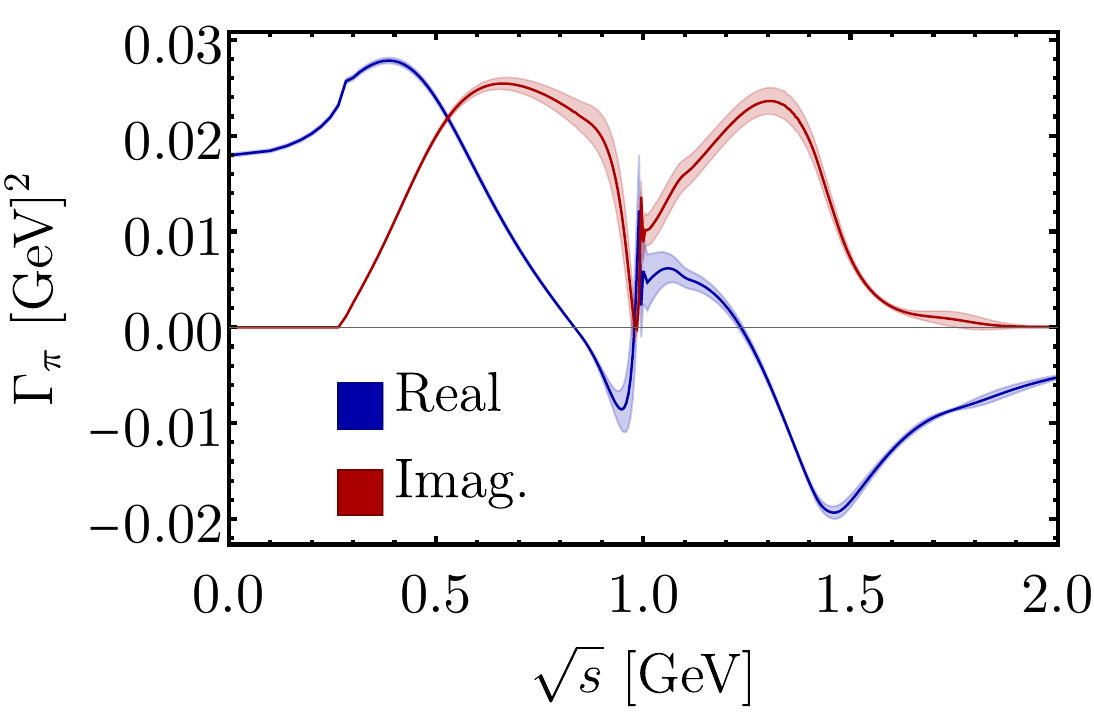}
         \caption{}
         \label{subfig:Gamma_pi}
     \end{subfigure}
     \hfill
     \begin{subfigure}{0.49\textwidth}
         \centering
         \includegraphics[width=\textwidth]{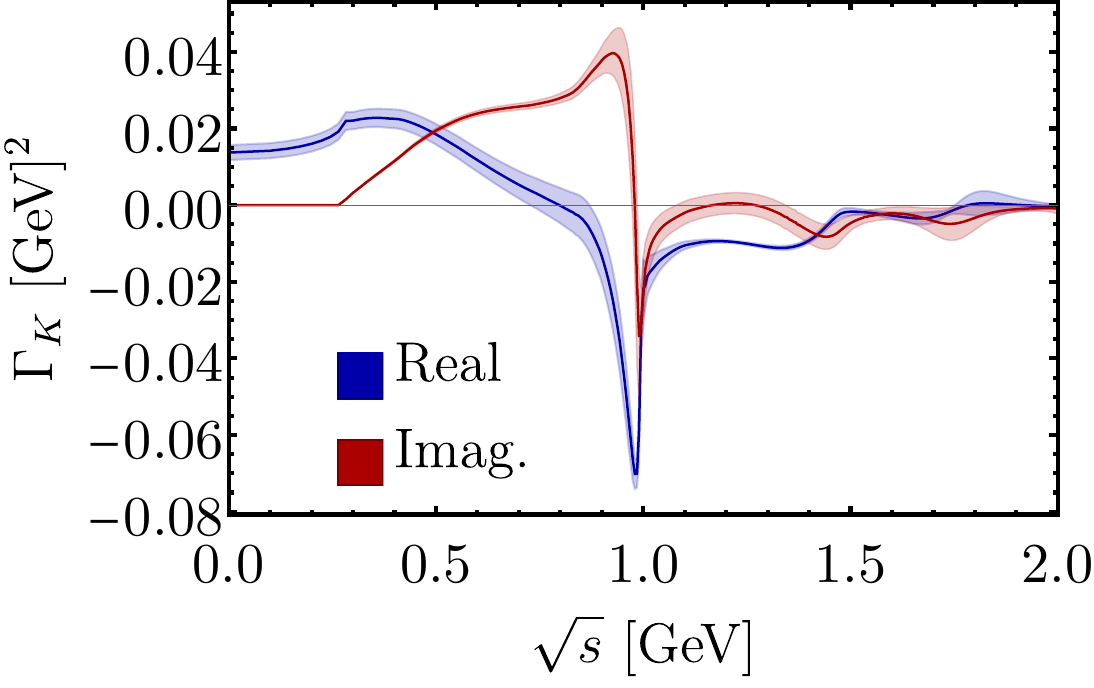}
         \caption{}
         \label{subfig:Gamma_K}
     \end{subfigure} \\
     
     \begin{subfigure}{0.49\textwidth}
         \centering
         \includegraphics[width=\textwidth]{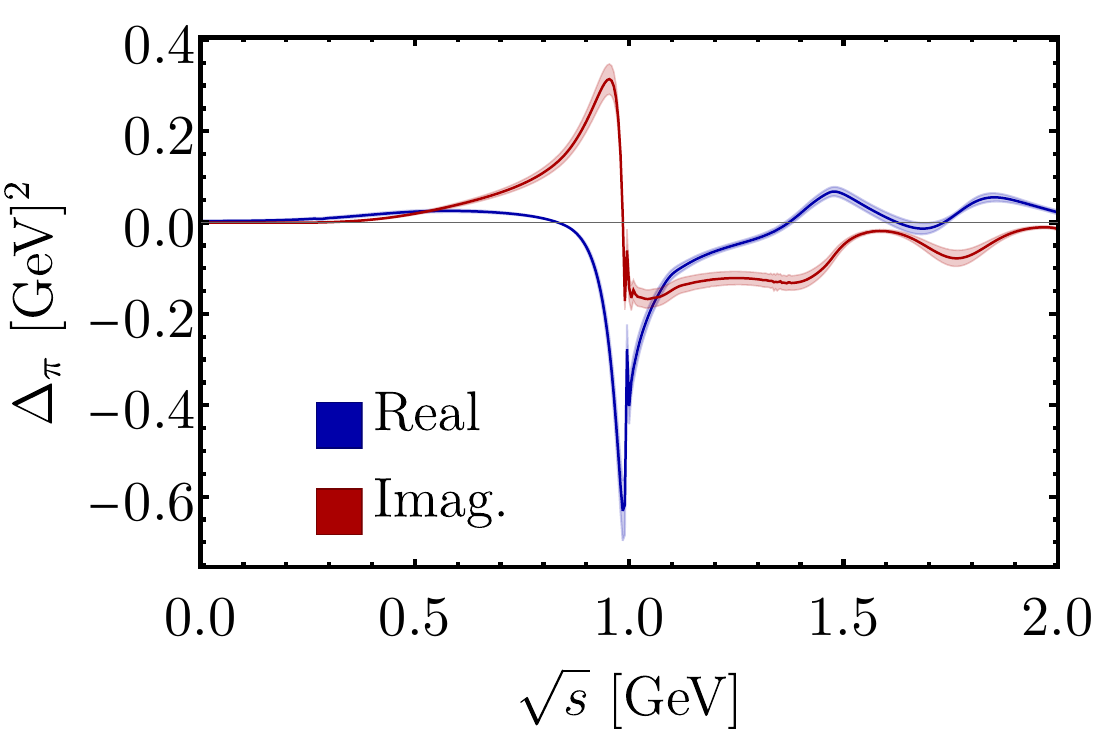}
         \caption{}
         \label{subfig:Delta_pi}
     \end{subfigure}
     \hfill
     \begin{subfigure}{0.49\textwidth}
         \centering
         \includegraphics[width=\textwidth]{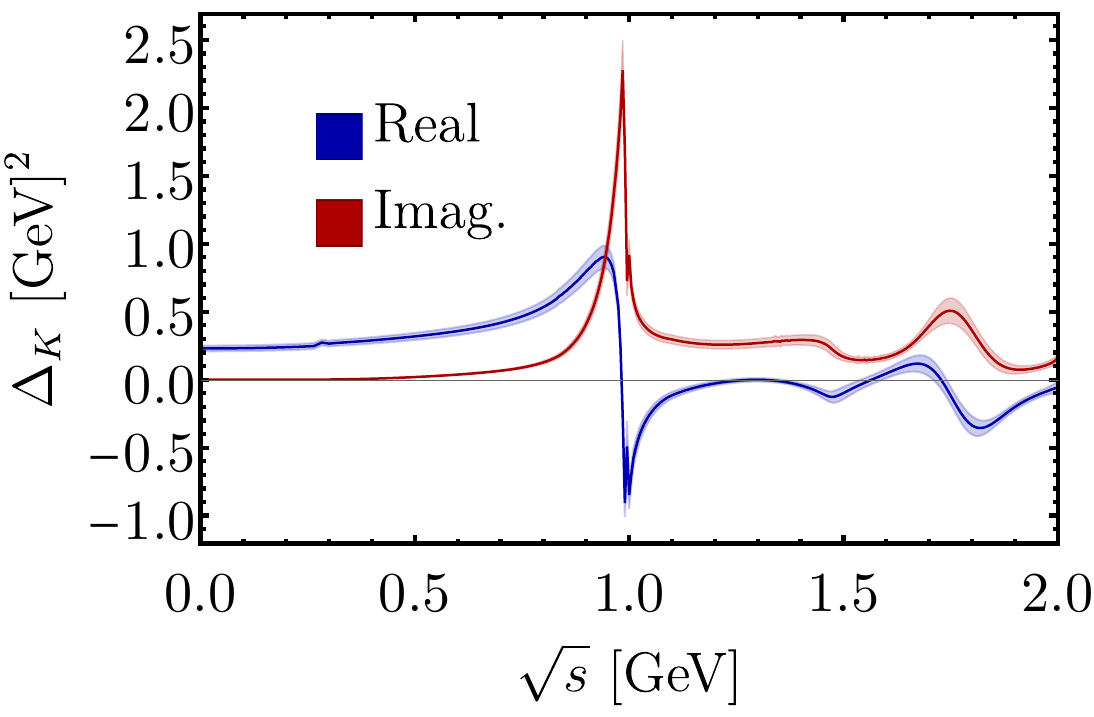}
         \caption{}
         \label{subfig:Delta_K}
     \end{subfigure} \\
 
    \begin{subfigure}{0.49\textwidth}
         \centering
         \includegraphics[width=\textwidth]{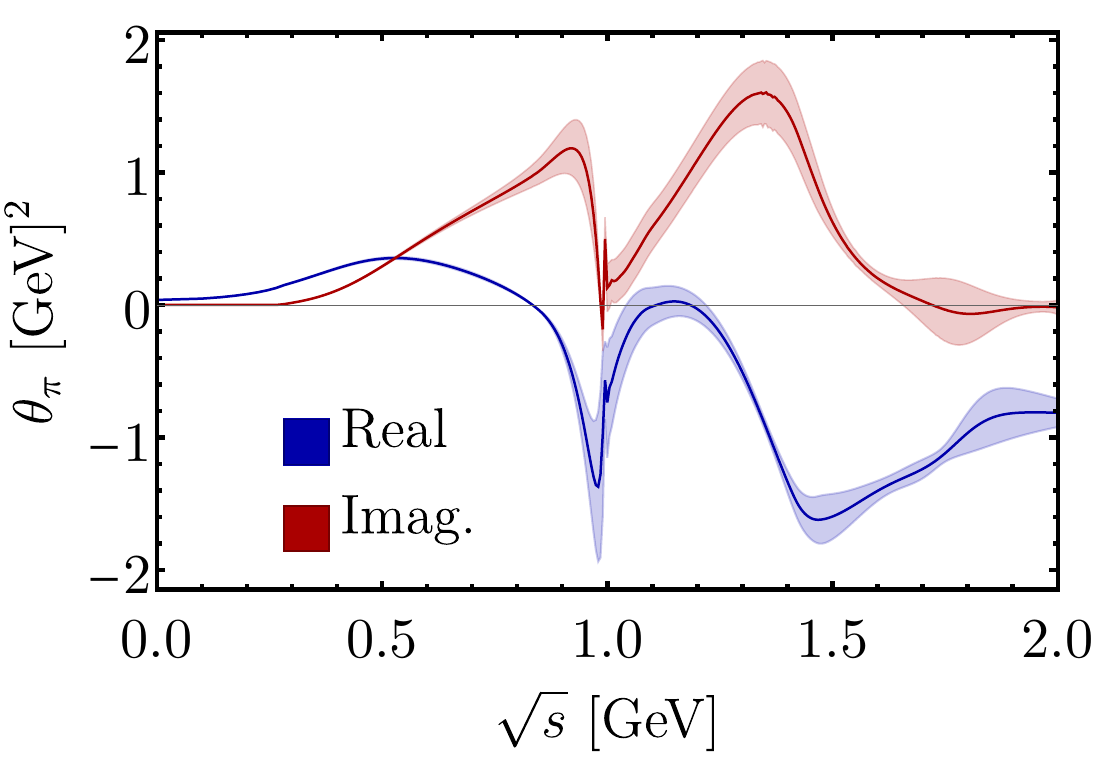}
         \caption{}
         \label{subfig:theta_pi}
     \end{subfigure} 
     \hfill 
     \begin{subfigure}{0.49\textwidth}
         \centering
         \includegraphics[width=\textwidth]{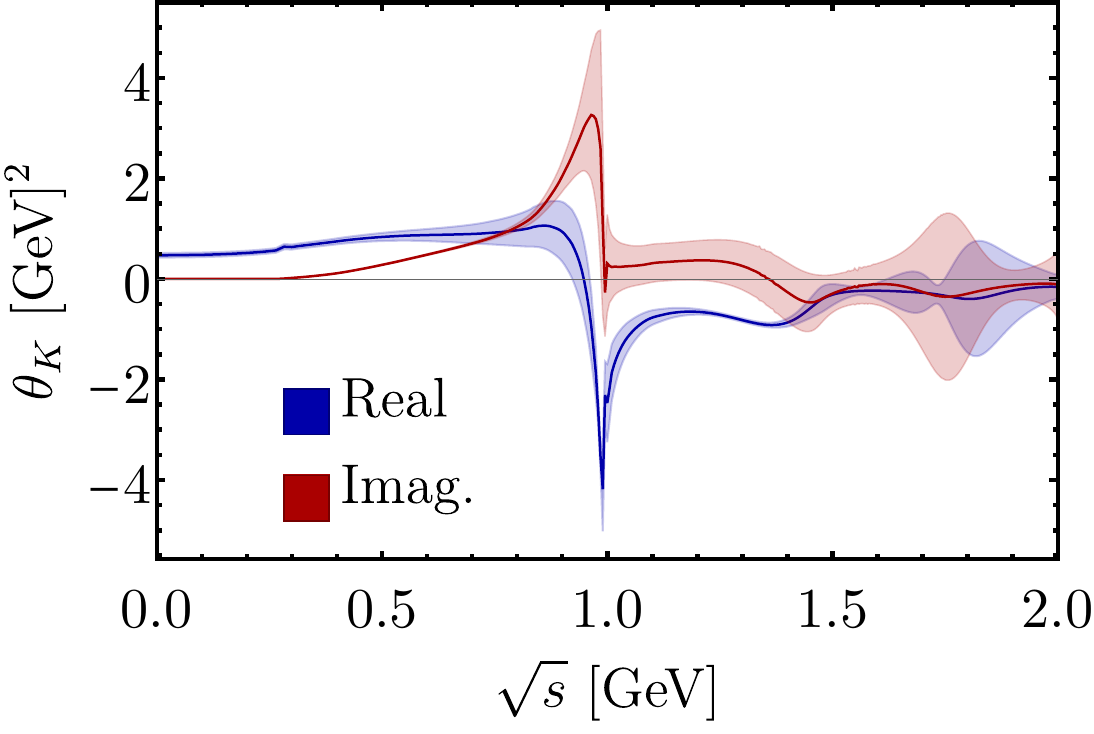}
         \caption{}
         \label{subfig:theta_K}
     \end{subfigure} 
     \caption{Plots of the real and imaginary parts of the form factors, $\Gamma_P$, $\Delta_P$, $\theta_P$, investigated in this work: (a),(c),(e) $F_\pi$ for $F \in \{\Gamma, \Delta, \theta\}$ respectively; (b),(d),(f) the same for $F_K$. The central, dark line is the mean of the Monte Carlo analysis with colored error bands reflecting the combined uncertainties from the Omnès functions parameter and low-energy matching parameters.}
     \label{fig:ffs_w_uncertainty}
\end{figure}

\begin{figure}[ht!]
     \centering
     \begin{subfigure}{0.49\textwidth}
         \centering
         \includegraphics[width=\textwidth]{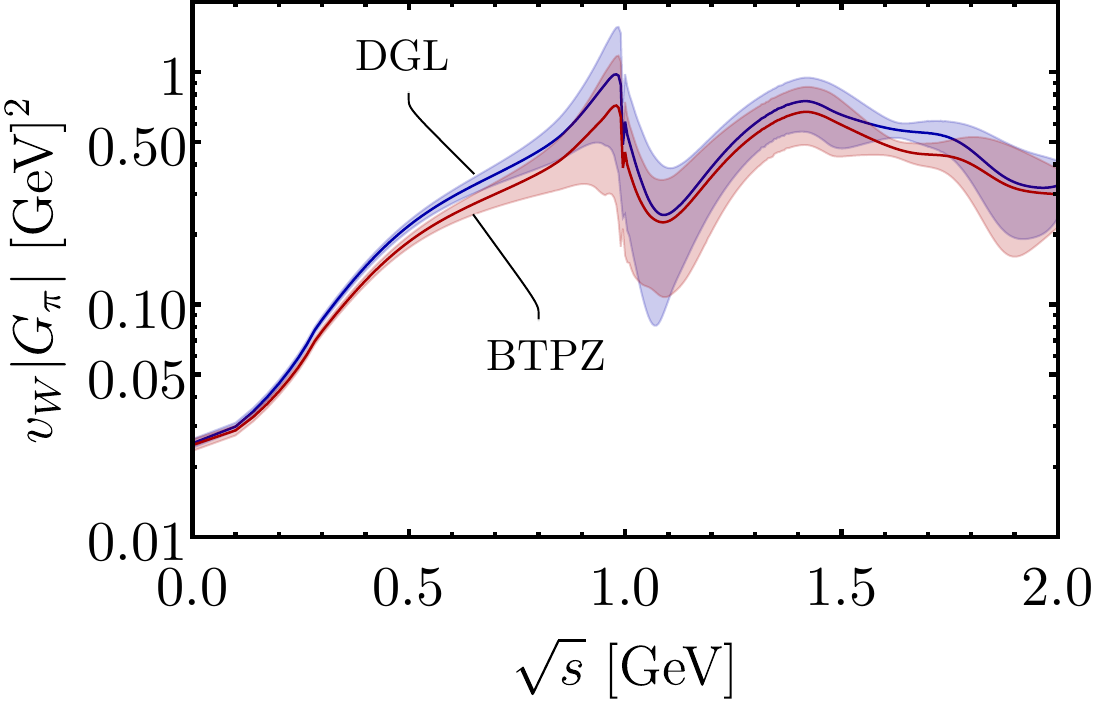}
         \caption{}
         \label{subfig:Gpi_dual}
     \end{subfigure}
     \hfill
     \begin{subfigure}{0.49\textwidth}
         \centering
         \includegraphics[width=\textwidth]{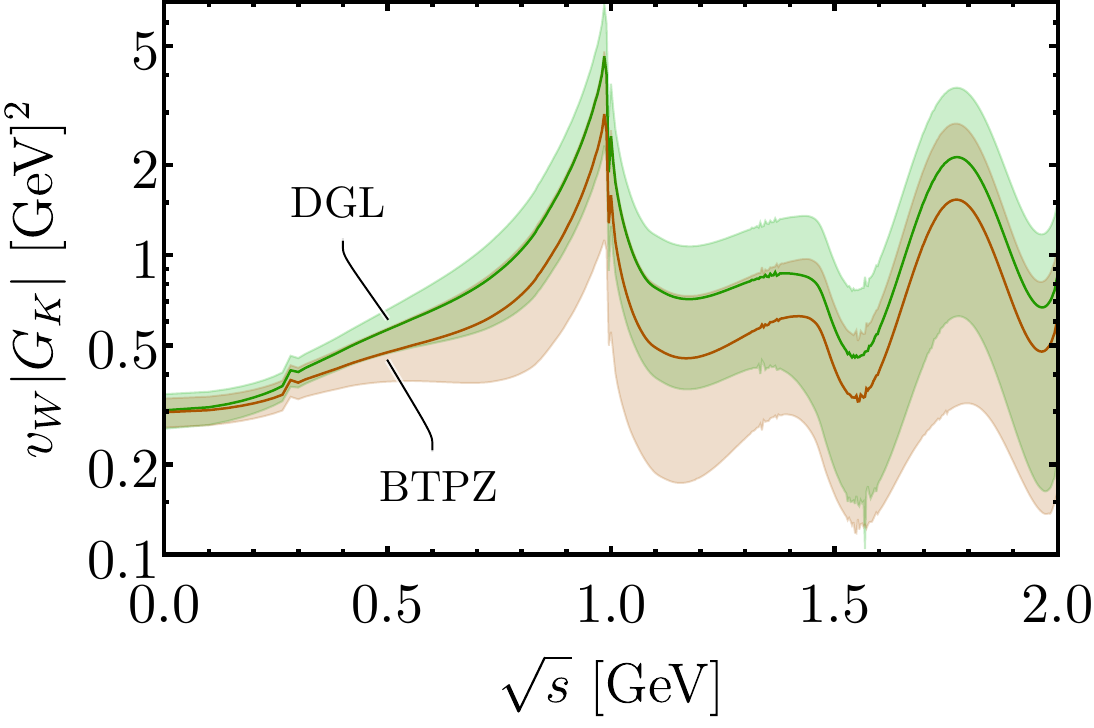}
         \caption{}
         \label{subfig:GK_dual}
     \end{subfigure}
     
        \caption{
            The (moduli of the) $G_P$ functions in the light Higgs-mixed scalar scenario with $\{c_{ud}, c_s, c_g\} = \{1,1,1\}$ for the two different low-energy matching conditions, derived (Eqs.~\eqref{s4e15} and \eqref{s4e16}) and direct (Eqs.~\eqref{s4e17} and \eqref{s4e18}). These figures illustrate the qualitative insensitivity of the physical results to the low-energy matching ambiguity. DGL (Donoghue, Gasser, \& Leutwyler) refers to \cite{Donoghue:1990xh} while BPTZ are the results of the current work. 
        }
    	\label{fig:Gs_with_uncertainty}
\end{figure}

Our results for $\Gamma_P, \Delta_P$, and $\theta_P$ form factors, including their uncertainty bands, are shown in Fig.~\ref{fig:ffs_w_uncertainty}. The form factor for the scalar decaying into pairs of pions or kaons, $G_P$, is shown in  Fig.~\ref{fig:Gs_with_uncertainty}. We plot $G_P$ for both sets of low-energy matching conditions and for the particular vectors of coupling constants, $\{c_{ud}, c_s, c_g\} = \{1,1,1\}$, corresponding to the scalar Higgs-mixed scenario (Eq.~\eqref{eq:higgs_mixed_condition}). We randomly sample the Omnès functions and the matching coefficients. The latter are assumed to follow a Gaussian distribution with mean and standard deviation corresponding to their main value and uncertainty as discussed in Section~\ref{sec:uncertainty_le_matching}. The solid lines and bands in Figs.~\ref{fig:ffs_w_uncertainty} and \ref{fig:Gs_with_uncertainty} correspond to the mean and one standard deviation, respectively, of this sample set. The two low-energy matching profiles overlap under this uncertainty prescription, which indicates that the order-of-magnitude conclusions are insensitive to this ambiguity. Due to the high degree of approximation, some numerical instability occurs at the division boundaries as $D_{\mathbf X}$ approaches a root, showing up in the plots as jagged, oscillatory behavior.

\subsection{Scalar decay widths}
\begin{figure}[ht!]
     \centering
     \begin{subfigure}{0.49\textwidth}
         \centering
         \includegraphics[width=\textwidth]{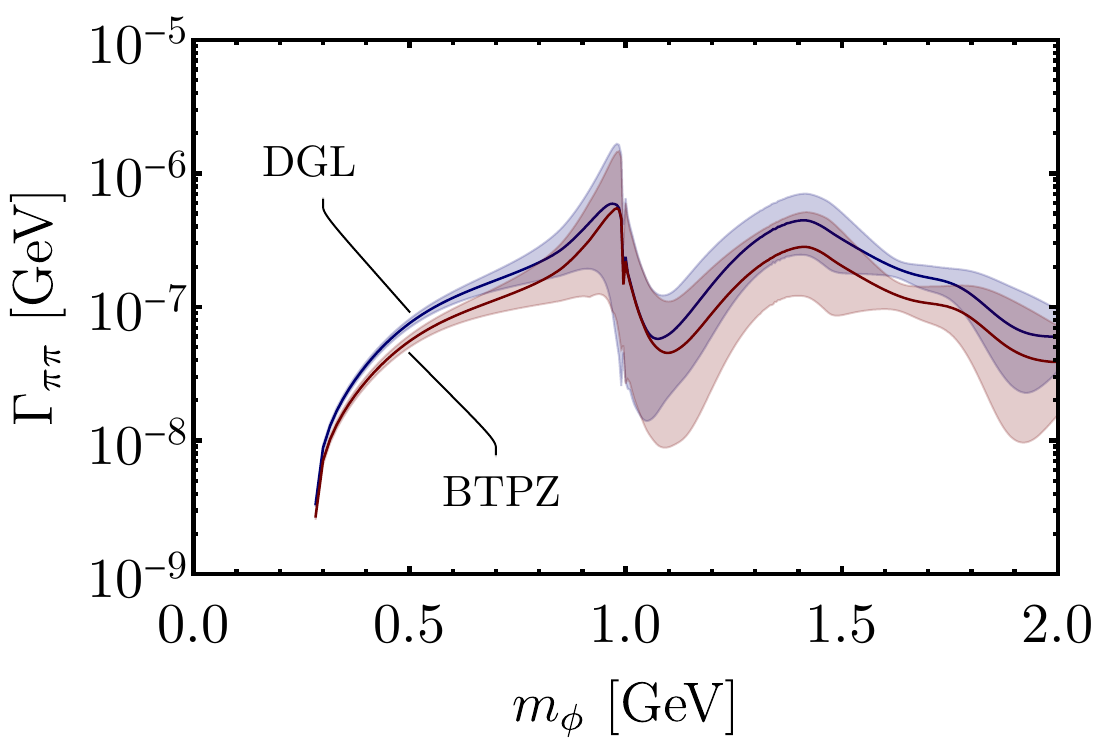}
         \caption{}
         \label{subfig:width_pipi_dual}
     \end{subfigure}
     \hfill
     \begin{subfigure}{0.49\textwidth}
         \centering
         \includegraphics[width=\textwidth]{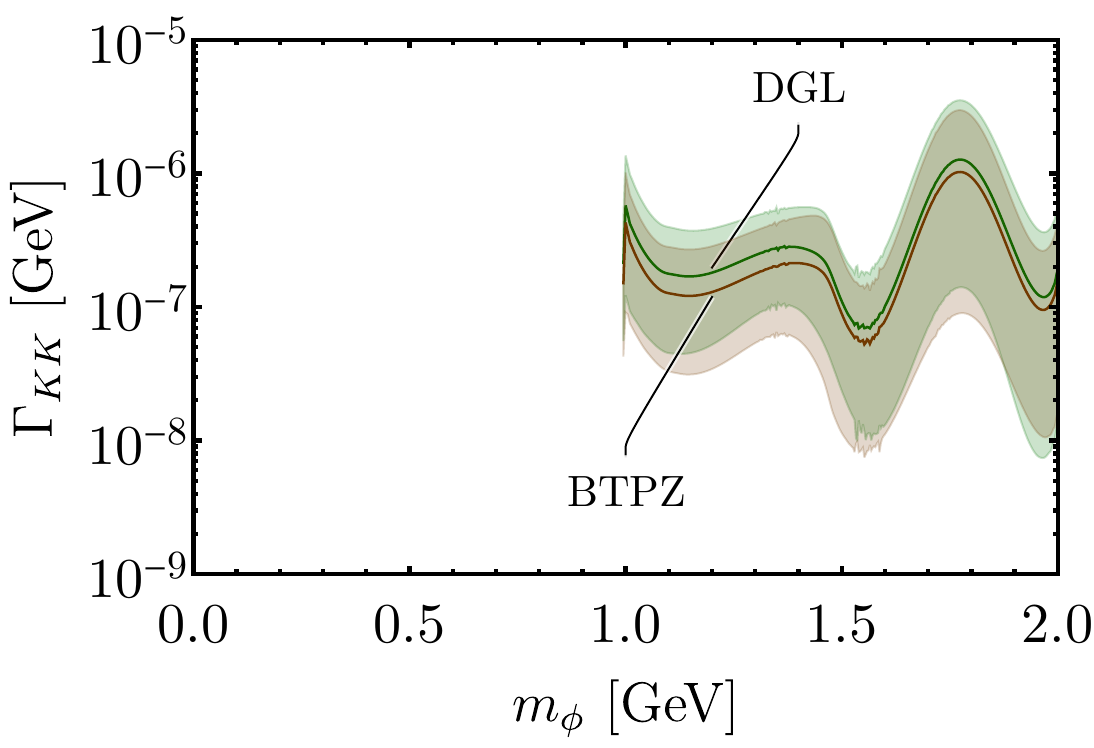}
         \caption{}
         \label{subfig:width_KK_dual}
     \end{subfigure}
        \caption{
            Decay widths $\phi \to PP$ for both candidate low-energy matching conditions. As in Figure~\ref{fig:Gs_with_uncertainty}, the larger of the two profiles in each plot (where distinguishable) are from the derived low-energy matching condition. The uncertainty bands in this plot are from the evaluation of the contours in Figure~\ref{fig:Gs_with_uncertainty}. 
        }
    	\label{fig:phi_to_PP_widths}
\end{figure}

We can compute the decay width of the light scalar $\phi$ into two pions or kaons using the form factors $G_P$, defined in Eq.~\eqref{s4e6}. The decay width reads as
\begin{align}
    \Gamma_{\phi \to PP} &= 
    \frac{ A_P }{ 16 \pi m_\phi } \sigma_P(m_\phi^2) |G_P(m_\phi^2)|^2 
    \label{eq:width_general}
\end{align}
where the number of final states is encoded by $A_\pi = 3$ ($\pi^\pm \pi^\mp$ and $\pi^0\pi^0$) and $A_K = 4$ ($K^\pm K^\mp$, $K^0 \overline{K}^0$ and $\overline{K}^0 K^0$).
Fig.~\ref{fig:phi_to_PP_widths} shows the widths $\Gamma \to PP$ for both direct and derived matching conditions. The three profiles (central, upper, and lower ridges) in each figure are obtained by plugging the corresponding profiles of Fig.~\ref{fig:Gs_with_uncertainty} into Eq.~\ref{eq:width_general}. 

\begin{figure}[ht!]
    \centering
    \includegraphics[width=\textwidth]{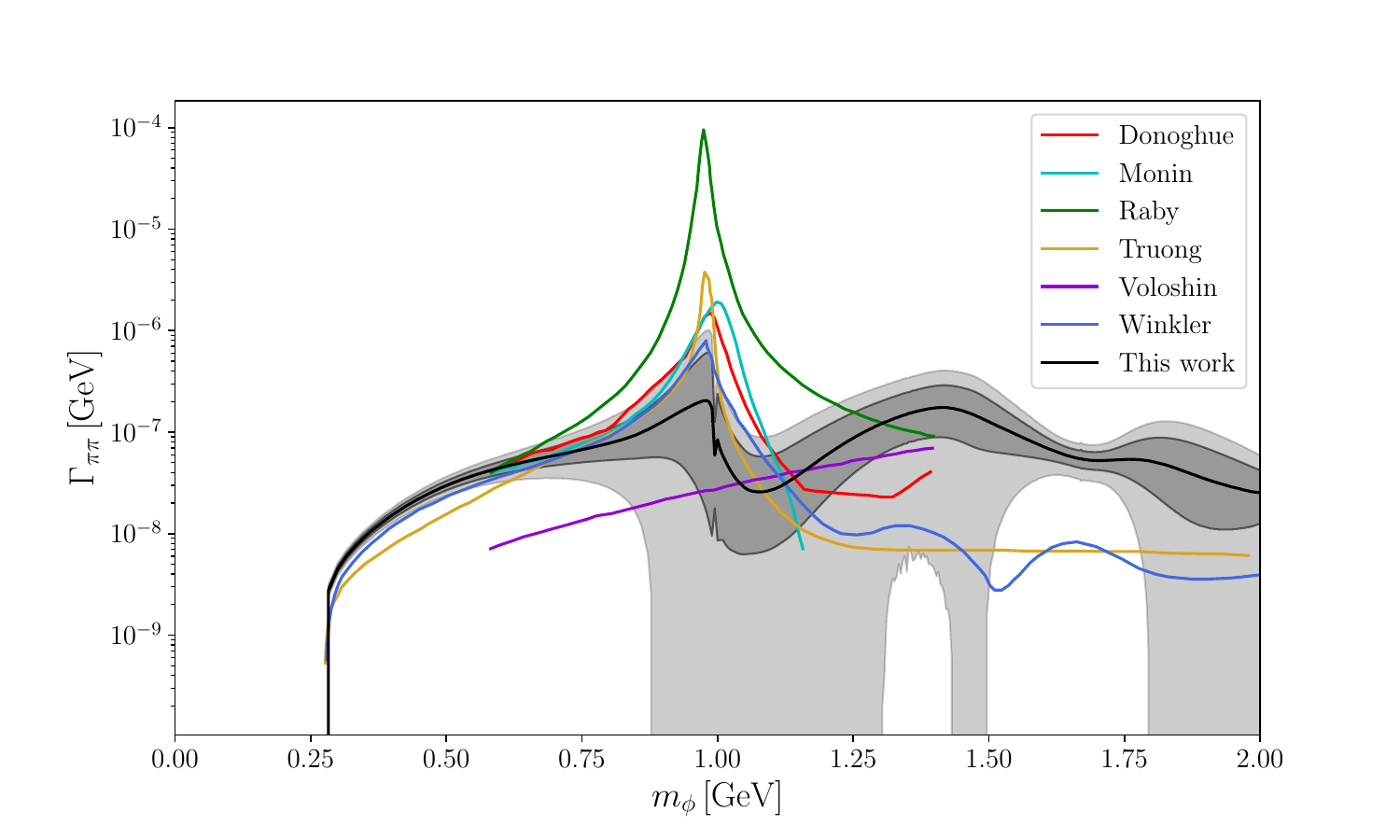}
    \caption{Comparison of our results with various previous works (Donoghue, Gasser \& Leutwyler \cite{Donoghue:1990xh}, Monin, Boyarsky \& Ruchayskiy \cite{Monin:2018lee}, Raby and West \cite{Raby:1988qf}, Truong \& Willey \cite{Truong:1989my}, Voloshin \cite{Voloshin:1985tc}, and Winkler \cite{Winkler:2018qyg}), as extracted from Figures 1 and 3 of \cite{Winkler:2018qyg}. Note that the Truong \& Willey profile is that with the sign in the $T$-matrix corrected. The direct method of low-energy matching has been employed for the contour from this work, and it is against the result of \cite{Winkler:2018qyg} that our contour is most directly comparable. In this comparison, $c_{ud}=c_{s}=c_{g}=s_\theta=1$. The dark gray band refers to the uncertainty band of Fig.~\ref{fig:phi_to_PP_widths}, while the light gray is twice this band's width to better illustrate the uncertainty on this logarithmic scale.}
    \label{fig:comparisons}
\end{figure}

In Fig.~\ref{fig:comparisons} we compare between our direct matching result for the $\phi \to \pi\pi$ decay width and those illustrated in Figs.~1 and~2 of \cite{Winkler:2018qyg}. We find good agreement up to $m_\phi \sim 0.75 \, \gev$ with most of the previous works. The peak generated by the opening of two kaon threshold is less pronounced than in other works, however the difference with the most recent ones~\cite{Donoghue:1990xh,Winkler:2018qyg} is attributable to the difference between derived and direct matching schemes and is inside uncertainty bands. From $m_\phi \sim 1.3 \, \gev$ onwards our results are consistently larger than those of Refs.~\cite{Truong:1989my,Winkler:2018qyg}.

\begin{figure}[ht!]
     \centering
     \begin{subfigure}{0.49\textwidth}
         \centering
         \includegraphics[width=1\textwidth]{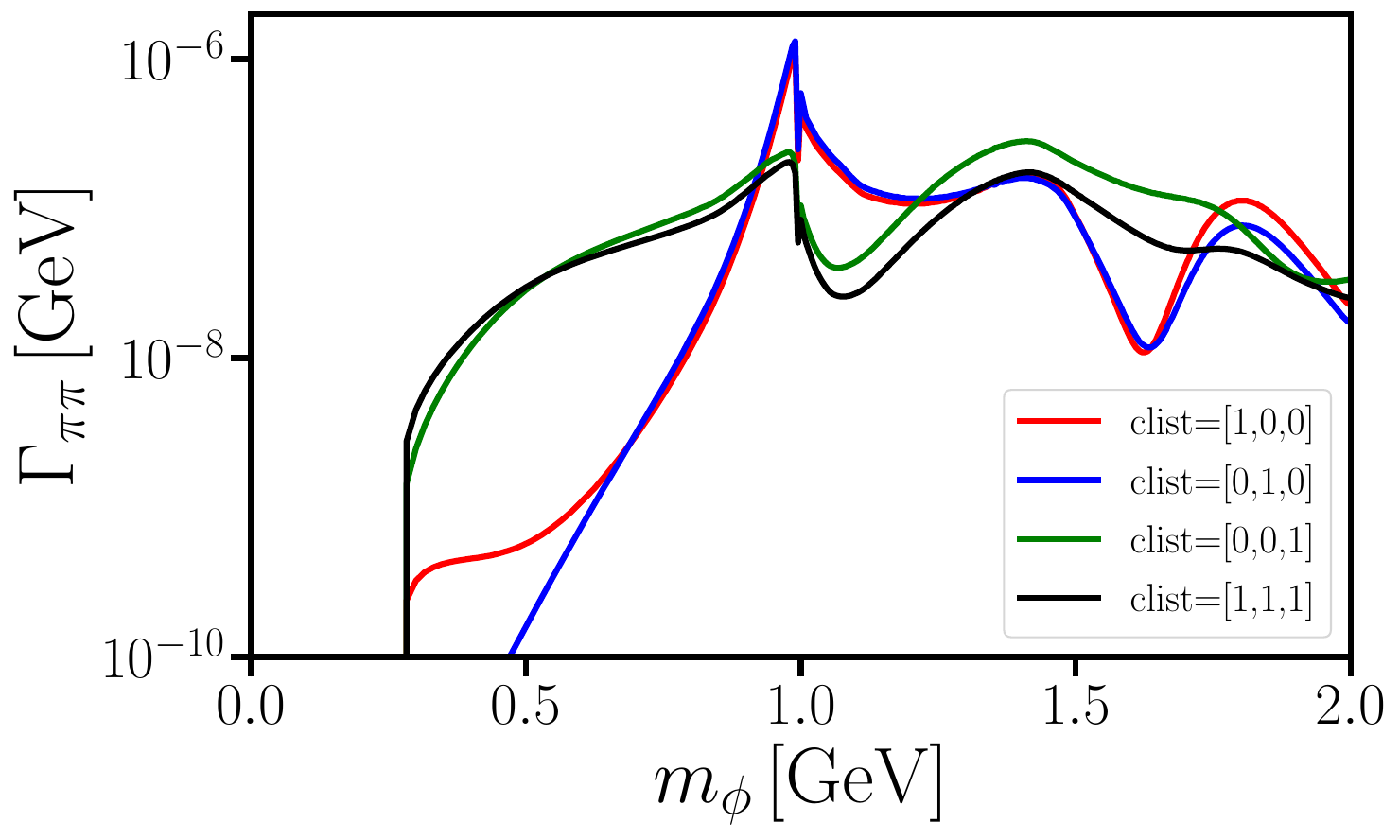}
         \caption{}
         \label{subfig:clist_comparison_pi}
     \end{subfigure}
     \hfill
     \begin{subfigure}{0.49\textwidth}
         \centering
         \includegraphics[width=1\textwidth]{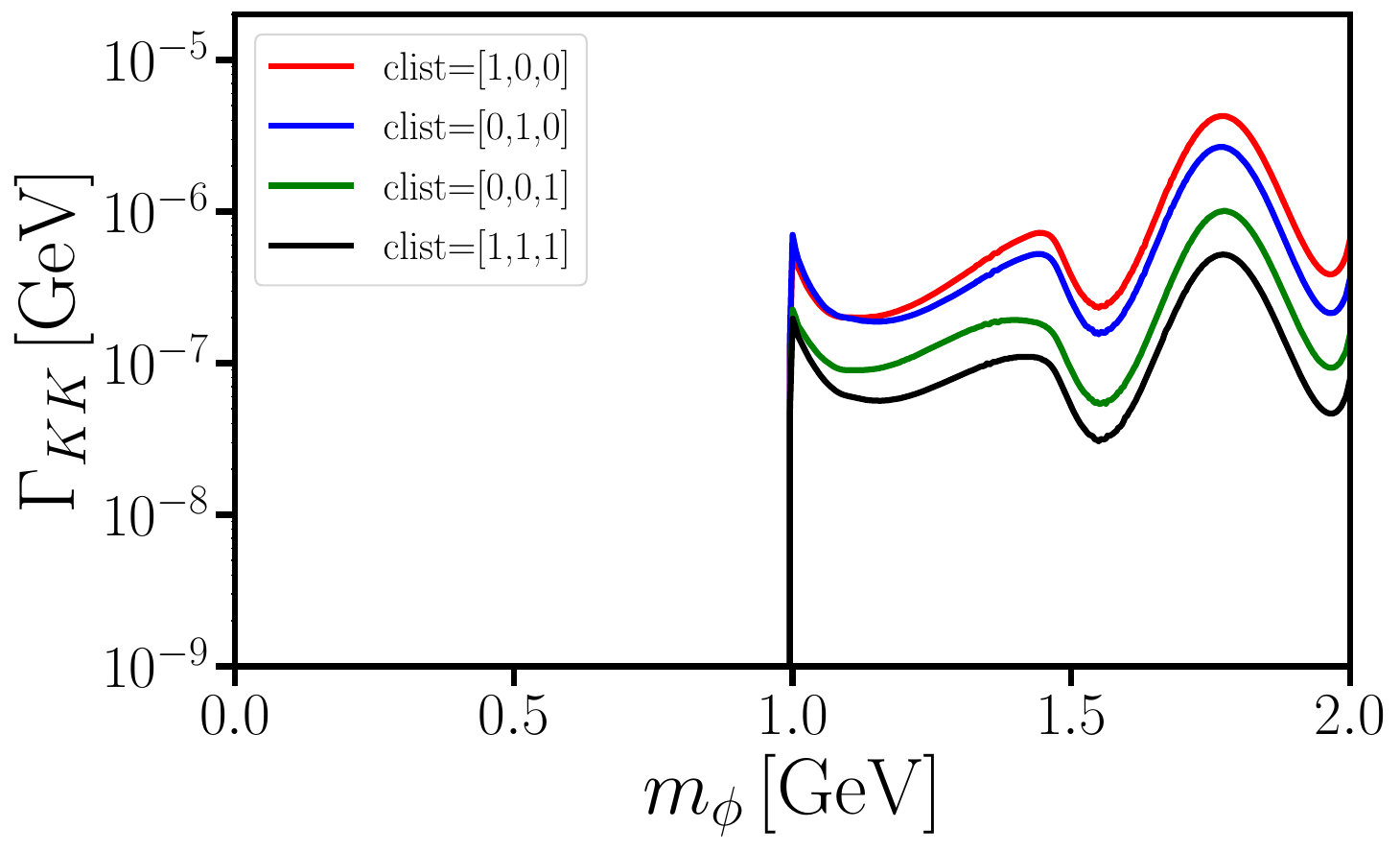}
         \caption{}
         \label{subfig:clist_comparison_K}
     \end{subfigure}
        \caption{Decay widths $\phi \to PP$ in the typical Higgs-mixed scenario plotted alongside the same expression with only a single of each of the three coupling constants from Eq.~\eqref{eq:lagrangian_general} turned on. These plots are an example of use of the public code presented in Appendix~\ref{app:code}. Note that, {\tt clist} corresponds to the parameter tuple $(c_{ud},c_s,c_g)$.
        }
    	\label{fig:clist_comparisons}
\end{figure}

The calculation of these decay widths for arbitrarily chosen couplings $c_{ud}, c_{s}, c_{g}$ can be computed using our results of the Omn\`es functions via the public code presented in Appendix~\ref{app:code}. We further exemplify the usage of the code's output in Fig.~\ref{fig:clist_comparisons}, where we explore the decay widths for specific choices of nonzero coupling constants of Eq.~\eqref{eq:lagrangian_general}. One noteworthy feature, for example, from Fig.~\ref{subfig:clist_comparison_pi}, is that the decay $\phi \to \pi\pi$ is seen to be primarily due to the coupling to the gluon term in Eq.~\eqref{eq:lagrangian_general}, while the effects of coupling to $c_{ud}$ and $c_s$ provide roughly the same effect beyond $m_\phi \gtrsim 0.75 \, \gev$. Meanwhile, this clear distinction is absent in the decay to kaons, Fig.~\ref{subfig:clist_comparison_K}.

%% file: 6_conclusion.tex
\section{Conclusion}\label{sec:conc}

We have reexamined the decays of a beyond the SM light scalar into pairs of pions and kaons. To do so, we have built a dispersive representation for these decays' form factors, which we denote as $G_P(s)$, $P=\pi,\,K$, for any values of the couplings of the light scalar with SM fields. The general solution for the dispersive representation, in Eq.~\eqref{gen_sol}, depends on a matrix of Omnès functions, which only depend on the nature of the low-energy intermediate states that generate the imaginary part of the form factors, in our case the coupled pion and kaon rescattering, and a set of subtraction polynomials that depend on the couplings of the light scalar with the pion and kaon pairs.

We have obtained the Omnès matrix function by solving Muskhelishvili-Omnès singular integral equations following the discretization technique proposed in Refs.~\cite{Moussallam:1999aq,Descotes-Genon:2000pfd}. These equations require as an input the $S$-wave isoscalar $\pi\pi\to \pi\pi$ and $\pi\pi\to K\bar{K}$ phase shifts, which we take from the parametrizations fitted to experimental data from Refs.~\cite{Garcia-Martin:2011iqs,Pelaez:2020gnd}. We propagate the uncertainty of these parametrizations, which reflects the uncertainty of the experimental data, into our results for the Omnès functions. We do so by obtaining a solution of a Gaussian sample of the parametrizations. The results for the Omnès functions including the uncertainties can be found in Fig.~\ref{fig:FFs_with_uncertainty}. Moreover, we explain in detail all the steps we have followed to obtain the numerical solutions.

The subtraction polynomials are obtained by matching the dispersive representation of the form factor to one valid for small $s$. The latter is obtained by splitting the form factor into three pieces corresponding to rewriting the SM current that couples with the light scalar in terms of the trace energy-momentum tensor, the $q=u,d$ mass, and the $s$ mass operators. These currents are commonly labeled as $\theta_P(s)$, $\Gamma_P(s)$ and $\Delta_P(s)$, respectively. If we consider the currents separately, the $\Gamma_P(s)$ and $\Delta_P(s)$ subtraction polynomials are just constants fixed by the normalization at $s=0$, which can be obtained from the Feynman-Hellmann theorem with the pion and kaon mass expressions obtained from $\chi$PT. In the case of $\theta_P(s)$ the overall normalization is fixed by the scale anomaly, however the  $\chi$PT representation depends on $s$ already at leading order. Thus, the subtraction polynomial must be of degree one and two conditions are needed to fix it: the normalization and the slope at $s=0$. The total uncertainty is obtained by combining the uncertainties of the Omnès functions and the one of the subtraction polynomial. The latter is reduced from previous determinations by using the NLO $\chi$PT expressions of the pion and kaon masses. Our results, including uncertainties, are shown in Fig.~\ref{fig:ffs_w_uncertainty}.

The determination of the subtraction polynomial for $G_P(s)$ presents an additional difficulty. If one obtains the subtraction polynomial by adding up the dispersive representations for $\theta_P(s)$, $\Gamma_P(s)$ and $\Delta_P(s)$ with the corresponding factors, as done in Ref.~\cite{Donoghue:1990xh}, one obtains a different solution from the one that one finds from building the dispersive representation of the full current $G_P(s)$. We call the first matching procedure \textbf{DGL} and the second \textbf{BTPZ}. Rewriting  $G_P(s)$ in terms of the trace of the energy momentum tensor introduces a dependence on $\beta$ and $\gamma_m$, for which we use  ${\mathcal O}(\alpha_s^4)$ expressions in order to minimize uncertainties. We show the results for both matchings in Fig.~\ref{fig:Gs_with_uncertainty}. The main sources of uncertainty are the determinations of $\dot{\theta}_\pi(0)$ and, in particular, of $\dot{\theta}_K(0)$, which are only known at LO in $\chi$PT.

Our results for the form factors, including uncertainty bands, can be found in Fig.~\ref{fig:ffs_w_uncertainty} for $\Gamma_P$, $\Delta_P$, $\theta_P$ and in Fig.~\ref{fig:Gs_with_uncertainty} for $G_P$. For the latter we show the results for the two matching procedures. Using $G_P(s)$, the decay widths for $\phi \to PP$ can be computed. We show our results with uncertainty bands using the BTPZ matching procedure for the specific set of couplings corresponding to the Higgs-mixed scalar scenario in Fig.~\ref{fig:phi_to_PP_widths}. Our predictions for the $\phi \to \pi\pi$ width is compared with previous works in Fig.~\ref{fig:comparisons}. To compute the widths and to produce plots for any set of couplings $(c_{ud}, c_s, c_g)$ we provide a public code, which can be downloaded and used following the instructions in Appendix~\ref{app:code}. A sample plot for several coupling sets is shown in Fig.~\ref{fig:clist_comparisons}.
\\
\\
{\bf Acknowledgements.} 
This work emerged from discussions at the Aspen Center for Physics, which is supported by National Science Foundation grant PHY-1607611. We would like to thank Jared Evans for providing the motivation for this study. We gratefully thank Heiri Leutwyler for very helpful observations and comments. EP warmly acknowledges insightful discussions with Bachir Moussallam.
This work was supported in part by the U.S. National Science Foundation Grant No. PHY-2310149 and by the U.S. Department of Energy contract DE-AC05-06OR23177. PB gratefully acknowledges a IU College of Arts and Sciences Dissertation Research Fellowship for completing this work. EP is supported in part by the Generalitat Valenciana (Spain) through the plan GenT program (CIDEGENT/2021/037). 
JZ acknowledges support in part by the DOE grant DE-SC1019775, and the NSF grant OAC-2103889. JZ acknowledges support in part by the Miller Institute for Basic Research in Science, University of California Berkeley.

%% file: app_ff_expressions.tex
\section{Public code}
\label{app:code}
The {\tt python} code associated with this publication can be found at \url{https://github.com/blackstonep/hipsofcobra} and installed from {\tt PyPI} by 
\begin{lstlisting}[language=Python]
python3 -m pip3 install hipsofcobra --user
\end{lstlisting}
The primary goal of the code is to compute the widths $\Gamma( \phi \to \pi\pi)$ and $\Gamma( \phi \to KK ) $ for a set of $c_{ud}, c_s$, and $c_g$ values input by the user (for the definitions see Eqs.~\eqref{eq:lagrangian_general} and \eqref{eq:cud}). The GitHub includes a Jupyter notebook ({\tt demos.ipynb}) which demonstrates how this is utilized by the end user, which we explicate here. 

The focus of the code is the {\tt HipsofCobra} class, which may be called as
\begin{lstlisting}[language=Python]
from hipsofcobra import HipsofCobra
import numpy as np
import matplotlib.pyplot as plt

hips = HipsofCobra( clist=[1,1,1], Pname='pi', method='DGL' ) 
\end{lstlisting}
where {\tt matplotlib} is necessary, if any of the automatic plotting functions of the class are to be called. The parameters of this class object are the $c=(c_{ud}, c_s, c_g)$ values desired for the computation, the final-state scalar particles (accepting {\tt `pi'} or {\tt `K'}), and the method for low-energy matching conditions ({\tt `DGL'} or {\tt `BTPZ'}). The final functionalities of this class are: 
\begin{enumerate}
    \item to write the widths to a comma-separated value (CSV) file, including upper and lower values for the width as discussed in this publication; 
    \item to produce plots of the form factors, $|G|$, with uncertainty contours; 
    \item to produce plots of the width files that were written to CSV; 
    \item lastly, to produce a plot of the many iterations involved in the computation of the $G$ values. 
\end{enumerate}
This fourth goal is mainly meant for illustration of the variations in profiles for different iterations of calculating $G$. Each of these goals are exemplified in the code below, the outputs of which are saved to a folder {\tt ./results/clist=[1,1,1]}, or whatever the user's choice of {\tt clist} was when defining {\tt hips}. 

\begin{lstlisting}[language=Python]
hips.write_widths() # Write widths csv file.

hips.plot_G_contours(
  color='k', xlim=[0,4], ylim=None, PrintQ=True, ShowQ=True
) # Produce pdf plot of |G|.

hips.plot_width_contours(
  color='k', xlim=[0,2], ylim=[1e-9,1e-5], PrintQ=True, ShowQ=True
) # Produce pdf plot of width phi->PP.

hips.plot_sl(
  xlim=[0,4], ylim=None, PrintQ=True, ShowQ=True
) # Produce plot of |G|s from all iterations. 
\end{lstlisting}
 The notation {\tt sl} stands for ``superlist,'' referring to the format of the Omn\`es input files included in the package. 

 As an illustration of the functionality of this code, Fig.~\ref{fig:clist_comparisons} plots the profiles of the decay widths for the Higgs-mixed scalar case alongside the three profiles that each retain only a single nonzero coupling constant, $c_{ud}$, $c_s$, or $c_g$ from Eq.~\eqref{eq:lagrangian_general}.

\section{Expressions for $\Gamma_P(0)$ and $\Delta_P(0)$} 
\label{app:ff_expressions}

The expressions for the meson masses in terms of the meson masses in $SU(3)$ at NLO~\cite{Gasser:1984gg} and in the isospin limit are
\begin{align}
    m^2_P &= m^2_{P,2}+m^2_{P,4}+\mathcal{O}(p^6)\,,\quad P=\pi,\,K,\, \\
    m^2_{\pi,2} &= 2 B_0 \hat{m}\,,\\
    m^2_{K,2} &= B_0(\hat{m} + m_s )\,, \\
    m^2_{\pi,4}
    &= 2 B_0 \hat{m}\left\{ \mu_\pi - \frac{1}{3} \mu_\eta + \frac{16 B_0 }{F_0^2 } \left[\hat{m} (2L^r_8 - L^r_5) + (2\hat{m}+m_s)( 2L^r_6 - L^r_4 ) \right] \right\}\,,\\
    \begin{split}
    m^2_{K,4} &= B_0(\hat{m}+ m_s) \\
    & \hspace{0.5cm} \times \left\{ \frac{2}{3} \mu_\eta + \frac{8 B_0 }{ F_0^2 } \left[(\hat{m}+m_s)( 2 L^r_8 - L^r_5 ) + 2(2\hat{m} + m_s) ( 2L^r_6 - L^r_4 ) \right] \right\}\,,
    \end{split}
\end{align}
where $\hat{m} = (m_u + m_d)/2$  and 
\begin{align}
    \mu_P &= \frac{ m_{P,2}^2 }{ 32 \pi^2 F_0^2 } \log \frac{ m_{P,2}^2 }{ \mu^2 }\,,
\end{align}
with $\mu$ the renormalization scale. The values of the LEC are taken from Ref.~\cite{Dowdall:2013rya} at $\mu=547 \, {\rm MeV}$.

Using the Feynman-Hellman theorem we obtain $\Gamma_P(0)$ and $\Delta_P(0)$ from the derivatives of the meson masses at NLO. Writing our expressions in terms of the physical meson masses, we find
\begin{align}
\begin{split}
\Gamma_\pi(0) =&\, m^2_{\pi}\biggr(1 + \frac{m^2_{\pi}}{F_0^2}\biggr\{\frac{1}{32\pi^2}\left(\frac{8}{9}+\log\frac{ m^2_{\pi} }{ \mu^2 } - \frac{1}{9}\log\frac{ m_\eta^2 }{ \mu^2 } \right) 
\\
&\hspace{4.5cm}+ 8[(2L^r_8-L^r_5 )+2(2L^r_6-L^r_4)] \biggr\}\biggr)+\mathcal{O}(p^6)\,,
               \end{split}
               \\
\Delta_\pi(0) =& \frac{m_\pi^2}{F_0^2 }\left(m^2_K-\frac{1}{2}m^2_\pi\right) \left[ - \frac{1}{ 72 \pi^2 } \left( 1 + \log \frac{ m_\eta^2 }{ \mu^2 } \right) + 16 (2 L^r_6 - L^r_4) \right]+\mathcal{O}(p^6)\,,\\
\begin{split}
\Gamma_K(0) = &\, \frac{1}{2}m_{\pi}^2\left\{1+\mu_{\eta}-\mu_{\pi} +\frac{8}{F_0^2} \bigg[(2m_K^2-m_\pi^2) (2L^r_8 - L^r_5)+4m_K^2(2L^r_6 - L^r_4) \right.
 \\
&\hspace{5.5cm} \left. \left. + \frac{m^2_K}{72\pi^2} \left( 1+\log \frac{m_\eta^2}{\mu^2} \right) \right] \right\}+\mathcal{O}(p^6)\,,
\end{split}
\\
\begin{split}
\Delta_K(0) =&\,\left(m_K^2 - \frac{1}{2} m_\pi^2\right)\left(1+ \frac{m_K^2}{F_0^2}\left\{ \frac{1}{36\pi^2} \left( 1 + \log \frac{m_\eta^2}{\mu^2} \right) + 8[(2L^r_8-L^r_5) + 2(2L^r_6-L^r_4)] \right\} \right) 
\\
             &+ \frac{m_\pi^2}{2} \left(\mu_{\pi}-\mu_{\eta} -\frac{8}{F^2_0}(m_K^2-m_\pi^2)(2 L^r_8 - L^r_5) \right)+\mathcal{O}(p^6)\,.
             \end{split}
 \end{align}